\documentclass[%
 reprint,
superscriptaddress,
 amsmath,amssymb,
prb,
]{revtex4-1}

\usepackage{graphicx}
\usepackage{dcolumn}
\usepackage{bm}
\usepackage{float}

\usepackage{xtab,afterpage} 

\newcommand*{\citen}[1]{%
  \begingroup
    \romannumeral-`\x 
    \setcitestyle{numbers}%
    \cite{#1}%
  \endgroup   
}

\begin{document}

\preprint{}

\title{A coupled magneto-structural continuum model for multiferroic $\mathrm{BiFeO}_3$}%

\author{John Mangeri}
 \email[]{john.mangeri@list.lu}
 \affiliation{Materials Research and Technology Department, Luxembourg Institute of Science and Technology, \\
 5 Av. des Hauts-Fourneaux, 4362 Esch-sur-Alzette, Luxembourg}
  \author{Davi Rodrigues}%
\affiliation{Department of Electrical Engineering, Politecnico di Bari, Via Edoardo Orabona, 4, 70126 Bari BA, Italy}%
\author{Monica Graf}
 \affiliation{Materials Research and Technology Department, Luxembourg Institute of Science and Technology, \\
 5 Av. des Hauts-Fourneaux, 4362 Esch-sur-Alzette, Luxembourg}
 \author{Sudipta Biswas}
 \affiliation{Computational Mechanics Division, Idaho National Laboratory, Idaho Falls, ID, USA}
 \author{Olle Heinonen}%
 \thanks{Present and permanent address of O.H.: Seagate Technology, 7801 Computer Avenue, Bloomington MN 55435}
\affiliation{Materials Science Division, Argonne National Laboratory, 9700 S Cass Ave, Lemont, Illinois, USA
}%
\author{Jorge Iniguez}%
 \email[]{jorge.iniguez@list.lu}
 \affiliation{Materials Research and Technology Department, Luxembourg Institute of Science and Technology, \\
 5 Av. des Hauts-Fourneaux, 4362 Esch-sur-Alzette, Luxembourg}
\affiliation{Department of Physics, University of Luxembourg, Luxembourg}%

\date{\today}

\begin{abstract}
A continuum approach to study magnetoelectric multiferroic $\mathrm{BiFeO}_3$ (BFO) is proposed.
Our modeling effort marries the ferroelectric (FE) phase field method and micromagnetic simulations in order to describe the entire multiferroic order parameter sector (polarization, oxygen antiphase tilts, strain, and magnetism) self-consistently on the same time and length scale.
In this paper, we discuss our choice of ferroelectric and magnetic energy terms and demonstrate benchmarks against known behavior.
We parameterize the lowest order couplings of the structural distortions against previous predictions from density functional theory calculations giving access to simulations of the FE domain wall (DW) topology.
This allows us to estimate the energetic hierarchy and thicknesses of the numerous structural DWs.
We then extend the model to the canted antiferromagnetic order and demonstrate how the ferroelectric domain boundaries influence the resulting magnetic DWs.
We also highlight some capabilities of this model by providing two examples relevant for applications.
We demonstrate spin wave transmission through the multiferroic domain boundaries which identify rectification in qualitative agreement with recent experimental observations.
As a second example of application, we model fully-dynamical magnetoelectric switching, where we find a sensitivity on the Gilbert damping with respect to switching pathways.
We envision that this modeling effort will set the basis for further work on properties of arbitrary 3D nanostructures of BFO (and related multiferroics) at the mesoscale.
\end{abstract}

\maketitle


\section{Introduction}

The phenomenological description of ferroic phase transitions is characterized by the onset of one or more order parameters below a critical temperature.
In the case of ferroelectric materials, the order parameter is an electric dipole condensed from unstable phonon modes \cite{Scott1974, RabeBook2007}.
For ferromagnets, a net nonzero magnetization arises as ordering dominates thermal spin fluctuations below the Curie point \cite{StohrBook}.
In both cases, the theoretical portrayal of a single order parameter (and its conjugate electric or magnetic field) has been quite successful in illustrating and driving interest in a plethora of functional materials properties of technological relevance.
Multiferroics are compounds where multiple order parameters coexist and are coupled together in non-trivial ways.
Magnetoelectric (ME) multiferroics exhibit ferroelectricity along with a magnetic ordering (which can be ferromagnetic \cite{Kezsmarki2015}, antiferromagnetic \cite{Heron2014}, ferrimagnetic \cite{Lin2017}, helimagnetic \cite{Seki2009}, etc.).
In the context of applications for electronics, these types of structures are very promising since the coupling can provide a pathway to controlling the magnetic (electric) state with an electric (magnetic) field \cite{Heron2014, Fiebig2016, Spaldin2019}.
Or it is proposed that this coupling can give rise to new properties not present in either ferroelectric or magnetic state alone \cite{Fiebig2016}.
For most ME multiferroics however, this intrinsic coupling can be quite weak leading to an interest in searching for materials candidates where this is not the case.
A particular ME multiferroic, the perovskite $\mathrm{BiFeO}_3$ (BFO), has been demonstrated to host appreciable spin-orbit coupling between its ferroelectric (FE) and antiferromagnetic (AFM) ordering.
In bulk, BFO undergoes a phase transition to a rhombohedral ferroelectric phase upon cooling below 1100 K \cite{Moreau1971,Smith1968} along with a N\'{e}el temperature of around 640 K resulting in collinear G-type AFM order \cite{Moreau1971}.
%
Due to its high transition temperatures, it is a promising material for applications at ambient conditions.
In BFO, the polarization \textbf{P} displays an 8-fold symmetry of domain states aligned along the pseudocubic [111] or equivalent directions.
The rhombohedral polar distortion (displacement of the $\mathrm{Bi}^{3+}$ and $\mathrm{Fe}^{3+}$ atoms relative to the oxygen atoms) is also accompanied by a spontaneous antiphase tilting of the $\mathrm{FeO}_6$ octahedral oxygen cages about the polar axis.
As such, the presence of the antiphase tilts at adjacent iron sites underpin an antisymmetric Dzyaloshinskii-Moriya interaction (DMI) which causes a canting of the anti-aligned Fe spins \cite{Ederer2005, Dixit2015}.
Therefore, BFO displays a weak net ferromagnetic moment $\mathbf{M}$ due to \emph{noncollinearity} in its magnetic structure.
In many samples or in bulk, this canted moment forms a long-period cycloid with a period of around 64 nm \cite{Sando2013, Agbelele2017, Burns2020, Xu2021}.
%

Due to its exceptional properties, BFO has been proposed to be used in a number of novel device concepts including beyond-CMOS logic gates \cite{Manipatruni2019, Parsonet2022}, tunneling magnetoresistant spintronic valves \cite{Bea2006, Fusil2014, Yin2018}, THz radiation emitters \cite{Takahashi2006, Guzelturk2020}, enhanced piezoelectric elements \cite{Paull2022, Heo2022}, ultrafast acoustic modulators \cite{Lejman2014}, or linear electrooptical components \cite{Zhu2016,Sando2014}.
As miniaturization is a significant concern for next generation device proposals, the thicknesses of these ME films synthesized for the aforementioned applications are in the range of a few 10s of nm to a few $\mu$m's \cite{Burns2020}. 
As highlighted in recent work \cite{Gross2017,Chauleau2019}, the observed spin cycloid abruptly changes propagation direction at the FE domain walls (DWs) indicating its strong coupling to the polar order.
%
%
Local measurement techniques suggest that the $109^\circ$-$71^\circ$-$109^\circ$ sequence of FE DWs display a Bloch-like character with $\mathbf{P}$ rotating across the DW with some sense of chirality \cite{Chauleau2019, Fusil2022} leading to open questions as to the driving force of this phenomena as well as if the ME coupling can also yield chiral magnetic textures at these DWs. 
Additionally, there have been other experimental observations of  unexplained mesoscopic phenomena in BFO.
Piezoforce microscopy measurements have revealed metastable states in epitaxial thin films where instead of the 8-fold possibility of domain orientations, there are 12 which also display an appreciable population of charged domain boundaries which are controllable by electric field cycling\cite{Park2013}.
A sought-after property of ME multiferroics is the ability to deterministically switch the magnetization with electric fields \cite{Heron2014}.
Due to the time and length scales involved in the practical implementations of ME switching, the dynamics of the coupled polar-magnetic texture is unclear.
Supporting theory utilizing atomistic methods can become computationally intractable due to too many atoms in the simulation box or a difficulty of modeling real interfacial or time-dependent phenomena.
As such, these methodologies can be difficult to implement to investigate the aforementioned experimentally relevant scenarios.
In order to investigate the mesoscopic picture of ME multiferroics taking into account both the ferroelectric physics and the micromagnetic formalism to describe the AFM behavior \cite{Ivanov2005}, we are motivated to develop a continuum model of BFO and its nanostructures.
The goal is to coarse-grain the materials physics into a predictive capability for large length and time scales in a single calculation.
While the phase field method has been particularly useful in understanding the ferroelectric domain topology and its response to external stimulii in BFO\cite{Chen2008, Xue2021}, a natural forward progression is to extend this type of continuum modeling to the spins in the material with micromagnetic simulations \cite{Liao2020a, Liao2020b}.
This would give access to new information about the collective spin excitations in the presence of (and coupled to) the topological defects (for example its domain walls or the recently experimentally resolved solitons \cite{Govinden2022} in BFO).
To explore these questions in this work, we propose a coupled multiferroic continuum model that marries the well-known FE phase field and micromagnetism self-consistently on the same time and length-scale.
In Sections \ref{sec:latticeE} and \ref{sec:FEgov}, we report a comprehensive description of the relevant governing equations and energy terms for the lattice contribution.
We study the FE DWs in Section \ref{sec:DWparam} and establish our predictions of $\mathbf{P}$ order parameter profiles (including also the spontaneous octahedral tilt and strain fields) for a number of different low-energy DWs in BFO.
This allows us to parameterize the model-specific gradient coefficients by comparing to density functional theory (DFT) calculations \cite{Dieguez2013}. 
Good agreement is demonstrated with respect to the energy hierarchy of the different low-energy DWs.
We also report our model's predictions of Bloch rotational components, residual strain fields, and thicknesses of different DW types. 
In Sections \ref{sec:magE}, \ref{sec:magHo}, and \ref{sec:magDW} we expand the model to include the magnetic order. 
We simulate the magnetic ground states in the presence of homogeneous and inhomogeneous structural order building on the results from the previous section.
We evaluate the influence of different types of polar domain boundaries also yielding estimates of the DW thicknesses, topology, and energies of the magnetic texture.
Then in Section \ref{sec:appl}, we provide two illustrative examples of the capabilities of our simulations: (i) spin-wave transport through the multiferroic DW boundaries highlighting their rectifying nature; (ii) fully-coupled \emph{dynamical} switching of the magnetization order with a time-dependent electric field through the ME effect demonstrating a non-trivial sensitivity on physical parameters. 
While our model (and the examples provided) is certainly not exhaustive, we hope that this work will set the basis for further studies on properties of arbitrary 3D BFO nanostructures (and related multiferroics) at the continuum approximation of theory.
%

%
\section{Multiferroic continuum model}\label{sec:model}
We consider a zero temperature limit free energy density functional defined as a sum of Landau-type energy density from the structural distortions of the lattice ($f_\mathrm{latt}$), the magnetic energy density due to the spin subsystem ($f_\mathrm{sp}$) and the magnetostructural coupling ($f_\mathrm{MP}$) in single crystal BFO,
\begin{align}\label{eqn:en_sum}
    f &= f_\mathrm{latt}(\mathbf{P},\mathbf{A},\bm{\varepsilon}) + f_\mathrm{sp}(\mathbf{L},\mathbf{m}) + f_\mathrm{MP}(\mathbf{L},\mathbf{m}, \mathbf{P},\mathbf{A}),
\end{align}
where lower case $f$ denotes a free energy \emph{density}.
In our continuum description, we need some formal definitions of the order parameters. 
The electric polarization $\mathbf{P}$ is connected to the displacement of $\mathrm{Bi}^{3+}$ and $\mathrm{Fe}^{3+}$ atoms relative to the oxygen anions.
The vector $\mathbf{A}$ describes the rotations of the $\mathrm{FeO}_6$ cages where the antiphase correlation between adjacent unit cells is implicitly assumed.
%
%
The spontaneous homogeneous strain that arises below the phase transition is the rank two tensor $\bm{\varepsilon}$ with symmetric components $\varepsilon_{ij} = \varepsilon_{ji}$,
\begin{align}\label{eqn:strain}
    \varepsilon_{ij} &= \frac{1}{2} \left(\frac{\partial u_i}{\partial x_j} + \frac{\partial u_j}{\partial x_i}\right),
\end{align}
where the variable $u_i$ is the component of the elastic displacement vector $\mathbf{u}$ which is solved for in our problem setup.
For the spin system, BFO is an antiferromagnet with anti-aligned spins at first-neighboring Fe sites (G-type) leading to two distinct sublattices $\mathbf{m}_1$ and $\mathbf{m}_2$.
The quantity $\mathbf{L}$ is the AFM N\'{e}el vector which we define as $\mathbf{L} = (\mathbf{m}_1 - \mathbf{m}_2)/2$. 
Additionally, we have the total magnetic moment $\mathbf{m} = (\mathbf{m}_1 + \mathbf{m}_2)/2$ which accounts for the weak nonvanishing magnetization that arises due to the DMI.
The quantities $\mathbf{L}$ and $\mathbf{m}$ are constrained such that $|\mathbf{L}| + |\mathbf{m}| = 1$ with, in general, $|\mathbf{L}| \gg |\mathbf{m}|$ and $\mathbf{L}\cdot\mathbf{m} = 0$ reflecting the presence of a strong AFM coupling between the sublattices but with a weak noncollinearity in $\mathbf{m}_1$ and $\mathbf{m}_2$.
%
%
The total weak magnetization can be computed as $\mathbf{M} = M_s \mathbf{m}$ where $M_s$ is the saturation magnetization density of the Fe sublattice ($4.0 \mu$B/Fe)\cite{Weingart2012,Xu2019,Xu2022}.

\subsection{Lattice energy}\label{sec:latticeE}

We define the free energy density corresponding to the structural distortions of the lattice as $f_\mathrm{latt}$,
\begin{align}\label{eqn:FE_en}
    f_\mathrm{latt} &= f_P + f_A + f_{AP} + f_{P\varepsilon} \\ \nonumber
    &+ f_{A\varepsilon} + f_\varepsilon + f_{\nabla P} + f_{\nabla A}. 
\end{align}
The energy expansion of $f_P, f_A$ and $f_{AP}$ contains only the terms allowed by symmetry to the fourth order\cite{Fedorova2022},
\begin{align}\label{eqn:FE_bulk}
    f_\mathrm{P} &= A_P \left(P_x^2 + P_y^2 + P_z^2 \right) + B_P \left(P_x^2 +  P_y^2 + P_z^2\right)^2 \\ \nonumber
    &+ C_P \left(P_x^2 P_y^2 + P_y^2 P_z^2 + P_x^2 P_z^2\right), \\ \nonumber
   f_\mathrm{A} &= A_A \left(A_x^2 + A_y^2 + A_z^2 \right) + B_A \left(A_x^2 +  A_y^2 + A_z^2\right)^2 \\ \nonumber
    &+ C_A \left(A_x^2 A_y^2 + A_y^2 A_z^2 + A_x^2 A_z^2\right).
\end{align}
and
\begin{align}\label{eqn:FE_coupled}
    f_\mathrm{PA} &= B_{PA} \left(P_x^2 + P_y^2 + P_z^2\right)\left(A_x^2 + A_y^2 + A_z^2\right) \\ \nonumber
    &+ C_{PA} \left(P_x^2 A_x^2 + P_y^2 A_y^2 + P_z^2 A_z^2\right) \\ \nonumber
    &+ C'_{PA} \left(P_x P_y A_x A_y + P_y P_z A_y A_z + P_x P_z A_x A_z \right).
\end{align}
Additionally, the elastic, electrostrictive ($\mathbf{P}$-$\bm{\varepsilon}$), and rotostrictive ($\mathbf{A}$-$\bm{\varepsilon}$) energy is included as
\begin{align}
    &f_\varepsilon = \frac{1}{2} C_{11} \left(\varepsilon_{xx} + \varepsilon_{yy} + \varepsilon_{zz} \right) \\ \nonumber
    &+ C_{12} \left(\varepsilon_{xx} \varepsilon_{yy} + \varepsilon_{yy} \varepsilon_{zz} + \varepsilon_{xx} \varepsilon_{zz} \right) \\ \nonumber
    &+ \frac{1}{2} C_{44} \left(\varepsilon_{xy}^2 + \varepsilon_{yz}^2 + \varepsilon_{xz}^2 \right), \\ \nonumber
    &f_{P\varepsilon} = q_{11} \left(\varepsilon_{xx} P_x^2 + \varepsilon_{yy} P_y^2 + \varepsilon_{zz} P_z^2\right) \\ \nonumber
    &+ q_{12} \left[\varepsilon_{xx} \left(P_y^2\hspace*{-1pt} + \hspace*{-1pt}P_z^2\right) + \varepsilon_{yy} \left(P_x^2\hspace*{-1pt} + \hspace*{-1pt}P_z^2\right) + \varepsilon_{zz} \left(P_x^2\hspace*{-1pt} + \hspace*{-1pt}P_y^2\right) \right]\\ \nonumber
    &+ q_{44} \left(\varepsilon_{yz} P_y P_z + \varepsilon_{xz} P_z P_x + \varepsilon_{xy} P_x P_y\right),
\end{align}
and
\begin{align}\label{eqn:FE_roto}
    &f_{A\varepsilon} = r_{11} \left(\varepsilon_{xx} A_x^2 + \varepsilon_{yy} A_y^2 + \varepsilon_{zz} A_z^2\right)\\ \nonumber
    &+ r_{12} \left[\varepsilon_{xx} \left(A_y^2\hspace*{-1pt} + \hspace*{-1pt}A_z^2\right) + \varepsilon_{yy} \left(A_x^2\hspace*{-1pt} + \hspace*{-1pt}A_z^2\right) + \varepsilon_{zz} \left(A_x^2\hspace*{-1pt} +\hspace*{-1pt} A_y^2\right)\right]\\ \nonumber
    &+ r_{44} \left(\varepsilon_{yz} A_y A_z + \varepsilon_{xz} A_z A_x + \varepsilon_{xy} A_x A_y\right)
\end{align}
respectively. 
Finally, to evaluate inhomogeneous phases (i.e DWs), we include the lowest-order Lifshitz invariants \cite{Cao1990, Li2001, Hlinka2006} for the structural distortions to Eq.~(\ref{eqn:FE_en}),
%
\begin{align}\label{eqn:gradP}
&f_\mathrm{\nabla P} =\frac{G_{11}}{2}  \left( P_{x,x}^2 + P_{y,y}^2 + P_{z,z}^2 \right) \\ \nonumber
&+  G_{12}  \left(P_{x,x} P_{y,y} + P_{y,y} P_{z,z} + P_{x,x} P_{z,z} \right) \\ \nonumber
&+ \frac{G_{44}}{2} \left[\left(P_{x,y} + P_{y,x} \right)^2+ \left(P_{y,z} + P_{z,y} \right)^2 + \left(P_{x,z} + P_{z,x}\right)^2\right]\\ \nonumber
\end{align}
and
\begin{align}\label{eqn:gradA}
&f_\mathrm{\nabla A} =\frac{H_{11}}{2}   \left( A_{x,x}^2 + A_{y,y}^2 + A_{z,z}^2 \right) \\ \nonumber
&+  H_{12}  \left(A_{x,x} A_{y,y} + A_{y,y} A_{z,z} + A_{x,x} A_{z,z} \right) \\ \nonumber
&+ \frac{H_{44}}{2} \left[\left(A_{x,y} + A_{y,x} \right)^2+ \left(A_{y,z} + A_{z,y} \right)^2 + \left(A_{x,z} + A_{z,x}\right)^2\right]\\ \nonumber
\end{align}
%
for both the $\mathbf{P}$ and $\mathbf{A}$ order parameters respectively.
A comma in the subscript denotes a partial derivative with respect to the specified spatial directions.
The bulk homogeneous contribution to the energy (i.e. the terms \emph{not} involving $f_{\nabla P}$ and $f_{\nabla A}$) has been previously parameterized with DFT calculations\cite{Fedorova2022}; we refer the reader to this publication for the relevant coefficients.
%
%

%
However, in the case of the gradient energy, the set of coefficients $\{ G_{ij}$, $H_{ij} \}$ are difficult to obtain directly from DFT (see for example the approach outlined in Refs. [\citen{Stengel2013, Stengel2016, Dieguez2022}]) - so we employ a fitting procedure in Sec. \ref{sec:DWparam} to evaluate them.
We should emphasize that if a different bulk homogeneous phenomenological potential is used (i.e. Refs. [\citen{Karpinsky2017, Marton2017, Xue2021}]), then the gradient coefficients obtained would be different since they depend strongly on the energetics of the order parameters in the vicinity of the DW.

\subsection{Governing equations}\label{sec:FEgov}

To find the polar ground states, we evolve the coupled time dependent Landau-Ginzburg (TDLG) equations,
\begin{equation}\label{eqn:TDLG_P}
\begin{aligned}
\frac{\partial \mathbf{P}}{\partial t} &= - \Gamma_P \frac{\delta f_\mathrm{latt}}{\delta \mathbf{P}} \\
\end{aligned}
\end{equation}
and
\begin{equation}\label{eqn:TDLG_A}
\begin{aligned}
\frac{\partial \mathbf{A}}{\partial t} &= - \Gamma_A \frac{\delta f_\mathrm{latt}}{\delta \mathbf{A}} \\
\end{aligned}
\end{equation}
along with satisfying the stress-divergence equation for mechanical equilibrium,
\begin{align}\label{eqn:stress-div}
\sum\limits_{j = x,y,z}\frac{\partial \sigma_{ij}}{\partial x_j} = 0,
\end{align}
where $\sigma_{ij} = \sigma_{ji} = \partial f_\mathrm{latt} / \partial \varepsilon_{ij}$ is the elastic stress of the material.
%
%
We write the components of $\sigma_{ij}$ as 
\begin{align}\label{eqn:totalstress}
\sigma_{ij} = \sum\limits_{k,l = x,y,z} C_{ijkl} \left(\varepsilon_{kl} + \varepsilon_{kl}^\mathrm{eig}\right)
\end{align}
where $\varepsilon_{kl}$ is the elastic strain from Eq.~(\ref{eqn:strain}) and the eigenstrain is related to the spontaneous strain via,
\begin{align}\label{eqn:eigenstrain}
\varepsilon_{ij}^\mathrm{eig} = \sum\limits_{k,l = x,y,z} \left(Q_{ijkl} P_k P_l + R_{ijkl} A_k A_l\right),
\end{align}
where $Q_{ijkl}$ and $R_{ijkl}$ are the electrostrictive and rotostrictive coefficients.
These are related to our free energy density coefficients $q_{ijkl}$ and $r_{ijkl}$ as (in Voight notation),
\begin{equation}\label{eqn:eig1}
\begin{aligned}
Q_{11} &= \frac{1}{3} \left(\frac{2\left(q_{11} - q_{12}\right)}{C_{11} - C_{12}} + \frac{q_{11} + 2 q_{12}}{C_{11}+2 C_{12}}\right), \\ 
\end{aligned}
\end{equation}
\begin{equation}\label{eqn:eig2}
\begin{aligned}
Q_{12} &= \frac{1}{3}\left(-\frac{q_{11}-q_{12}}{C_{11}-C_{12}} + \frac{q_{11} + 2 q_{12}}{C_{11}+2 C_{12}}\right),\\
\end{aligned}
\end{equation}
and
\begin{equation}\label{eqn:eig3}
\begin{aligned}
Q_{44} &= \frac{q_{44}}{4 C_{44}},\\
\end{aligned}
\end{equation}
%
%
with similar definitions for the quantities involving $R_{ijkl}$.
We also investigate electrostatic phenomena in our model through the Poisson equation,
\begin{equation}\label{eqn:poissonE}
\begin{aligned}
    \epsilon_b \nabla^2 \Phi_\mathrm{E} &= \nabla \cdot \textbf{P}, \\
\end{aligned}
\end{equation}
where $\Phi_\mathrm{E}$ is the electrostatic potential which defines the electric field $\mathbf{E} = - \nabla \Phi_\mathrm{E}$ in the usual way.
The parameter $\epsilon_b = 30 \, \epsilon_0$ is the relative background dielectric constant \cite{Graf2015}.
Eq.~(\ref{eqn:poissonE}) is solved at every time step of the evolution of Eq.~(\ref{eqn:TDLG_P}) and (\ref{eqn:TDLG_A}).
In Sec. \ref{sec:DWparam} we are searching for the local minima due to the relaxation dynamics of Eq.~(\ref{eqn:TDLG_P}) and (\ref{eqn:TDLG_A}) and as such the time relaxation constants $\Gamma_P$ and $\Gamma_A$ are set to unity.

To enforce periodicity on the strain tensor components in our representative volume element that includes DWs, we separate the strain fields calculated from Eq.~(\ref{eqn:strain}) and (\ref{eqn:stress-div}) into homogeneous (global) and inhomogeneous (local) parts.
This is done utilizing the method formulated by Biswas and co-workers in Ref. [\citen{Biswas2020}] which relaxes the stress components along the periodic directions and thus allows corresponding deformation to occur.
Here, the homogeneous contribution of the total strain obeys the following integrated quantity at every time step of the relaxation,
\begin{equation}
\begin{aligned}\label{eqn:globalstrain}
\int\limits_V d^3\mathbf{r} \,\, \sigma_{ij}^\mathrm{total}  = 0,
\end{aligned}
\end{equation}
where $V$ is the volume of our simulation containing the DW profiles.
The total stress tensor, $\sigma_{ij}^\mathrm{total}$, is calculated from the sum of homogeneous, inhomogeneous, and eigenstrain components $\varepsilon_{ij}^\mathrm{total} = \varepsilon_{ij}^\mathrm{inhom} + \varepsilon_{ij}^\mathrm{hom} + \varepsilon_{ij}^\mathrm{eig}$ for all \emph{periodic} directions $i$ and corresponding periodic component $j$ at every time step of 
the simulation.

\subsection{Numerical implementation}\label{sec:num}

Equations (\ref{eqn:TDLG_P}), (\ref{eqn:TDLG_A}), (\ref{eqn:stress-div}), (\ref{eqn:poissonE}), and (\ref{eqn:globalstrain}) are cast into their weak formulation sufficient for the finite element analysis.
Our method uses linear Lagrange shape functions for the coupled variable system.
The finite element mesh spacing is selected to be $\Delta x \approx 0.1$ nm for all calculations in this work.
This small mesh spacing helps resolve the thin DWs in BFO to smoothness which are discussed extensively in Section \ref{sec:DWparam} and \ref{sec:magDW}.
We implement Newmark-beta time integration \cite{Newmark1959} with convergence between time steps achieved when the nonlinear residuals calculated during the Newton-Raphson iteration (with block Jacobi preconditioning) have been reduced by $10^{-8}$ relative tolerance.
%
%
%
If convergence is not obtained, we use adaptive time stepping with reduction factor of $0.5$.
The finite element method (FEM) implementation of this work is available within \textsc{Ferret} \cite{Mangeri2017} which is an add-on module for the open source Multiphysics Object Oriented Simulation Environment (MOOSE) framework \cite{permann2020moose}.
In the absence of order parameter gradients, the homogeneous FE states of $\mathbf{P}$ parallel to $\mathbf{A}$ which we denote as $\mathbf{P}\uparrow\uparrow\mathbf{A}$ can be obtained numerically.
To perform this calculation, we evolve Eq.~(\ref{eqn:TDLG_P}) and (\ref{eqn:TDLG_A}) simultaneously solving Eq.~(\ref{eqn:stress-div}) (at every time step) until the relative change in total volume integrated energy density $F$ between adjacent time steps is less than $5\times 10^{-7}$ eV/s.
The bulk potential predicts the spontaneous values of the order parameters upon minimization that are $P_s = |\mathbf{P}| = 0.945 \,\,\mathrm{C}/\mathrm{m}^2$  and $A_s = |\mathbf{A}| = 13.398^\circ$.
The spontaneous normal and shear strains that correspond to these values are $\varepsilon_{n} = \varepsilon_{ii} = 1.308\times10^{-2}$ and $\varepsilon_{s} = \varepsilon_{ij} = 2.95\times10^{-3}$ for $i \neq j$ in agreement with Ref. [\citen{Fedorova2022}].
The free energy density of the ground state given by Eq.~(\ref{eqn:FE_en}) is -15.5653 $\mathrm{eV}\cdot\mathrm{nm}^{-3}$.
The energy functional used also describes identical energy minima when $\mathbf{P}\uparrow\downarrow\mathbf{A}$ (which is equivalent to a $180^\circ$ phase reversal of the tilt field).
%
%
Since the rotostrictive strains defined in Eq.~(\ref{eqn:eigenstrain}) are invariant upon full reversal of $\mathbf{A}$, then these numbers are left unchanged.
In the next Section \ref{sec:DWparam} we evaluate the inhomogeneous textures of the DWs and parameterize the gradient coefficients $\{G_{ij}, H_{ij}\}$ used in our model.
\subsection{Calculation of gradient coefficients}\label{sec:DWparam}
In order to study the domain wall topology involving spatial variations of $\mathbf{P}$, $\mathbf{A}$, and strain, a good parameter set estimate of the gradient coefficients $(G_{11}, H_{11}, ...)$ of Eq.~(\ref{eqn:gradP}) and Eq.~(\ref{eqn:gradA}) is needed.
To achieve this, we consult DFT calculations reported by Di\'{e}guez and co-workers in Ref. [\citen{Dieguez2013}].
It was shown that an assortment of metastable states are allowed in BFO and that this zoology of different DW types forms an energy hierarchy. 
Due to electrostatic compatibility, this collection of states has specific requirements on the components of the order parameters that modulate across the domain boundary.
For example, the lowest energy configurations which we denote (see Table~\ref{tab:FEdw1}) as $2/1 (100)$ and $3/0 (110)$ are the $109^\circ$ and $180^\circ$ DWs respectively.
In this notation, it is indicated that, for the $2/1$ DW, two components of $\mathbf{P}$ and one component of $\mathbf{A}$ switch sign across the boundary whose plane normal is (100), whereas for the $3/0$ DW, $\mathbf{P}$ undergoes a full reversal where $\mathbf{A}$ is unchanged across the (110)-oriented boundary plane.
We label the pairs of the domains characterizing the DW as $\mathbf{P}^\mathrm{I}/\mathbf{A}^\mathrm{I}$ and $\mathbf{P}^\mathrm{II}/\mathbf{A}^\mathrm{II}$ in this table.
This determines which terms in Eq.~(\ref{eqn:gradP}) and (\ref{eqn:gradA}) are primary contributions to the DW energy.
This is particularly advantageous as it has allowed us to separate the computation of specific DWs in the analysis of fitting the gradient coefficients to the DFT results.

\begin{figure}\centering
\hspace*{-5pt}\includegraphics[scale=0.285]{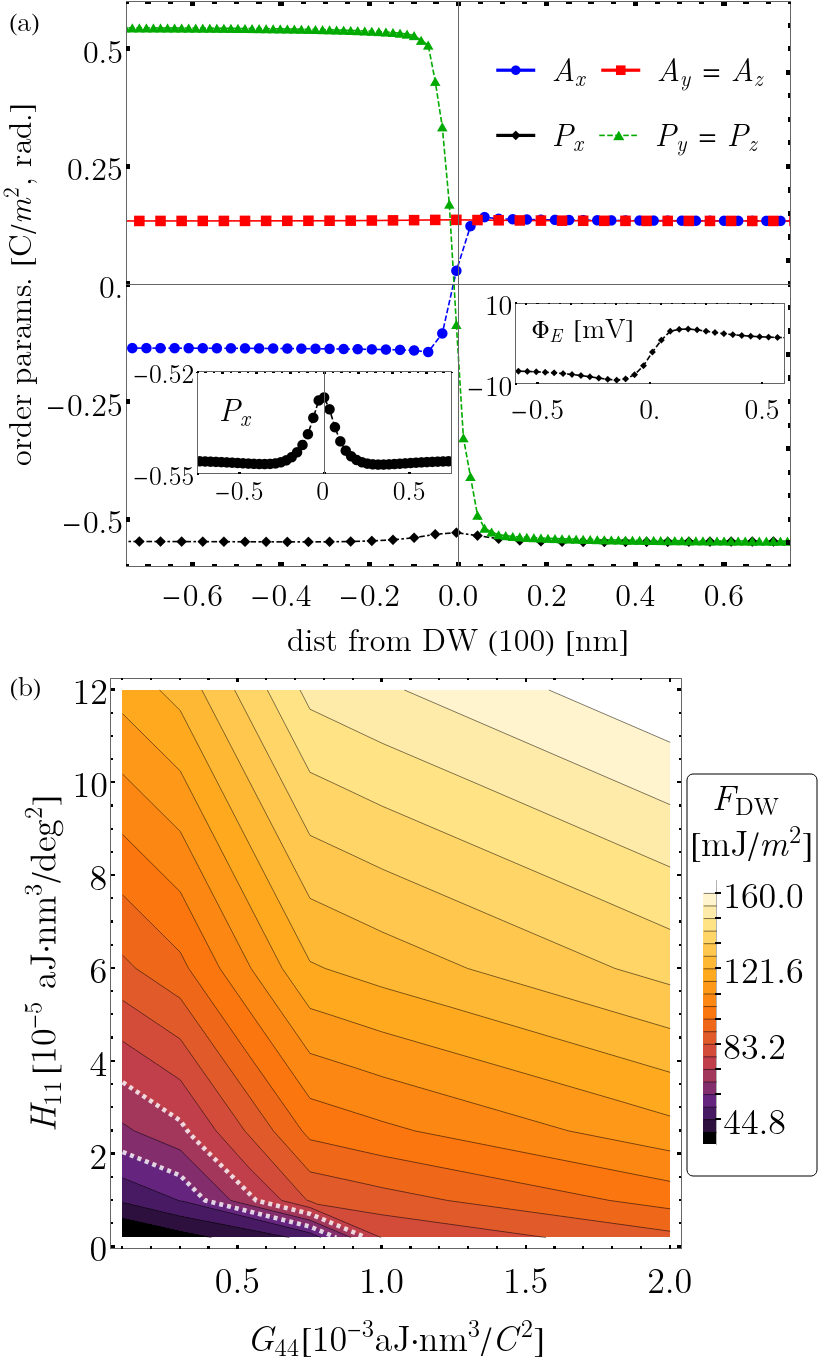}
\caption{\label{fig:21wall} (a) $\mathbf{P}$ and $\mathbf{A}$ 2/1 (100)-oriented DW profile. 
The left inset shows the $x$-component of $\mathbf{P}$ decrease across the DW where the right inset demonstrates the built-in $\Phi_\mathrm{E}$ (in mV) arising from this small rotation. (b) Energy surface as a function of primary gradient coefficients $H_{11}$ and $G_{44}$ with DW-DW distance of $\approx 160$ nm. For panels (a) the solution coincides with our best estimates of $G_{44}$ and $H_{11}$ (listed in Table \ref{tab:gradCoeff}).}
\end{figure}

To obtain the (100)- or (110)-oriented DWs within our phase field scheme, we choose an initial condition for the components of the order parameters to be a sin(x) or sin(x+y) profile respectively.
We then relax Eq.~(\ref{eqn:TDLG_P}), and (\ref{eqn:TDLG_A}) until convergence along with satisfying the conditions of mechanical equilibrium of Eq.~(\ref{eqn:stress-div}) at every time step.
The periodic boundary conditions on the components of $\mathbf{P},\mathbf{A},$ and $\mathbf{u}$ for (100)- or (110)-oriented domain walls are enforced along the [100] and [110] directions respectively.
We compute the DW energy with
\begin{equation}\label{eqn:enDW}
\begin{aligned}
    F_\mathrm{DW} &= \frac{F - F_0}{N\cdot S} \\
\end{aligned}
\end{equation}
where $F_0$ is the corresponding monodomain energy from Eq.~(\ref{eqn:FE_en}) integrated over the computational volume $V$. 
The energy $F$ is computed from the solution that contains the DW profile.
The number of DWs in the simulation box is $N$ and $S$ the surface area of the DW plane.
We find convergence on the computed energies within 1 $\mathrm{mJ}/\mathrm{m}^2$ provided that the DW-DW distances are greater than 30 nm due to long-range strain interactions.
For fourth-order thermodynamic potentials, a fit function of the form $W_k \tanh{\left[\left(r-r_0\right)/t_k\right]}$ is sufficient to fit the evolution of order parameters that switch across the DW\cite{Marton2010} where $W_k$ is the value of the switched spontaneous order parameters far from a DW plane localized at $r_0$ and $t_k$ corresponds to the thickness of the polar or octahedral tilt parameters for $k = P,A$ respectively.

\begin{table*}
\caption{\label{tab:FEdw1} Types of (100)- and (110)-oriented domain walls, their primary derivatives and corresponding gradient coefficients, and comparison of energies calculated from DFT\cite{Dieguez2013} with those in this work. Adjacent domain configurations for $\mathbf{P}$ and $\mathbf{A}$ utilize the $\mathrm{I}$ and $\mathrm{II}$ superscript notation as discussed in the main text. Energy is presented in $\mathrm{mJ}/\mathrm{m}^2$ and DW thicknesses ($2 t_k$, $k=P,A$) are given in nm.}
\begin{tabular*}{\textwidth}{c|cc|c|cccccccc|cc|cccc}
\hline\hline
 &  & & & & & & & & & & & & &  \\ 
$\textbf{P}^\mathrm{I}/\mathbf{A}^\mathrm{I}$ & Type & DW & $\textbf{P}^\mathrm{II}/\mathbf{A}^\mathrm{II}$ & & $P_{i,j}$ & & $G_{ij}$& & $A_{i,j}$ &  & $H_{ij}$ & $2 t_P$ & $2 t_A$ & & $F_\mathrm{DW}^{(\mathrm{DFT})}$ &  & $F_\mathrm{DW}^{(\mathrm{FEM})}$  \\ 
\colrule
 &  & & & & & & & & & & & &  \\ 
$[111]/[111]$ & $0/0$ & - & $[111]/[111]$ & & $-$ & & $-$ & & $-$ & & $-$ & $-$ & $-$ &  & $-$ & & $-$ \\ 
$[111]/[111]$ & $0/3$ & $(100)$ & $[111]/[\bar{1}\bar{1}\bar{1}]$ & & $-$ & & $-$ & & $A_{x,x}, A_{y,x}, A_{z,x}$ & & $H_{44}, H_{11}$ & $-$ & $0.39$ & & $227$ & & $293$  \\ 
$[111]/[111]$ & $1/1$ & $(100)$ & $[11\bar{1}]/[11\bar{1}]$ & & $P_{z,x}$ & & $G_{44}$ & & $A_{z,x}$ & & $H_{44}$ & $0.33$ & $0.52$ &  & $151$ & & $162$ \\ 
$[111]/[111]$ & $1/2$ & $(100)$ & $[11\bar{1}]/[\bar{1}\bar{1}1]$ & & $P_{x,x}$ & & $G_{11}$ & & $A_{y,x}, A_{z,x}$  & & $H_{44}$ & $0.25$ & $0.25$ & & $147$ & & $159$  \\ 
$[111]/[111]$ & $2/1$ & $(100)$ & $[1\bar{1}\bar{1}]/[\bar{1}11]$ & & $P_{y,x}, P_{z,x}$ & & $G_{44}$ & & $A_{x,x}$ & & $H_{11}$ &$0.08$ &$0.06$ & & $62$ & & $60$  \\
$[111]/[111]$ & $2/2$ & $(100)$ & $[1\bar{1}\bar{1}]/[1\bar{1}\bar{1}]$ & & $P_{y,x}, P_{z,x}$ & & $G_{44}$ & &  $A_{y,x}, A_{z,x}$ & & $H_{44}$ & $0.42$ & $0.34$ & & $319$ & & $314$ \\
$[1\bar{1}1]/[1\bar{1}1]$ & $3/0$ & $(110)$ & $[\bar{1}1\bar{1}]/[1\bar{1}1]$ & & $P_{x,x} P_{y,x}, P_{z,x}$ & & $G_{11}, G_{12},$ & &  $-$ & & $-$ & $0.28$& $-$ & & $74$ & & $78$ \\
 &  & & & & $P_{x,y} P_{y,y}, P_{z,y}$ & & $G_{44}$ & & & & & & &  \\
$[1\bar{1}1]/[1\bar{1}1]$ & $3/3$ & $(110)$ & $[\bar{1}1\bar{1}]/[\bar{1}1\bar{1}]$ & & $P_{x,x} P_{y,x}, P_{z,x}$ & & $G_{11}, G_{12},$ & &  $A_{x,x}, A_{y,x}, A_{z,x}$ & & $H_{11}, H_{12},$ & $0.22$& $0.33$ & & $255$ & & $263$ \\
 &  & & & & $P_{x,y} P_{y,y}, P_{z,y}$ & & $G_{44}$ & & $A_{x,y}, A_{y,y}, A_{z,y}$ & & $H_{44}$ & & &  \\
\hline\hline
\end{tabular*}
\end{table*}

As a first example, consider the lowest energy DW predicted by DFT, the so-called $109^\circ$ 2/1 (100) DW which is indeed frequently observed in thin film samples of BFO \cite{Zhang2018, Parsonet2022}. 
The primary gradient coefficients governing the energy of the wall are the $H_{11}$ and $G_{44}$ coefficients owing to the fact that $A_{x,x}$, $P_{y,x}$, and $P_{z,x}$ are nonzero (see Table \ref{tab:FEdw1}). 
The resulting DW profile for the 2/1 (100) wall is presented in Fig.~\ref{fig:21wall}(a).
The profile is a smooth rotation of both $A_x$ and $P_y = P_z$ across the wall region.
The inset on the left reveals that the \emph{non-switching} component $P_x$ experiences a slight decrease ($\approx -3\%$) at the wall.
The quantitative value of the modulation of the non-switched component is consistent with DFT results of the same DW type\cite{Korbel2020}.
The small change of $P_x$ corresponds with a built-in $\Phi_\mathrm{E}$ shown in the right inset panel which is of comparable order ($\approx 10$ mV) to those estimated from DFT\cite{Korbel2020}.
Fitting the $\mathbf{P}$-$\mathbf{A}$ profile shows that the DW is quite thin (thickness $2 t_P \approx 0.08$ nm). Hence, we obtain DWs with marked Ising character.
We provide an energy profile scan across the primary coefficients $H_{11}$ and $G_{44}$ in (b).
The dashed white line outlines the predictions from DFT results in Ref. [\citen{Dieguez2013}].
We also should mention that the dependence on other coefficients is quite weak due to the relatively small gradients in non-switching components.
These calculations (and others not shown here) reveal that the choice of $\{G_{ij},H_{ij}\}$ is not unique, i.e. one can find the same DW energies (with very similar profiles) for different combinations of the primary coefficients.
Therefore, it is necessary to visit other DW configurations to constrain the values of the entire set.

\begin{figure*}\centering
\hspace*{-12pt}\includegraphics[scale=0.258]{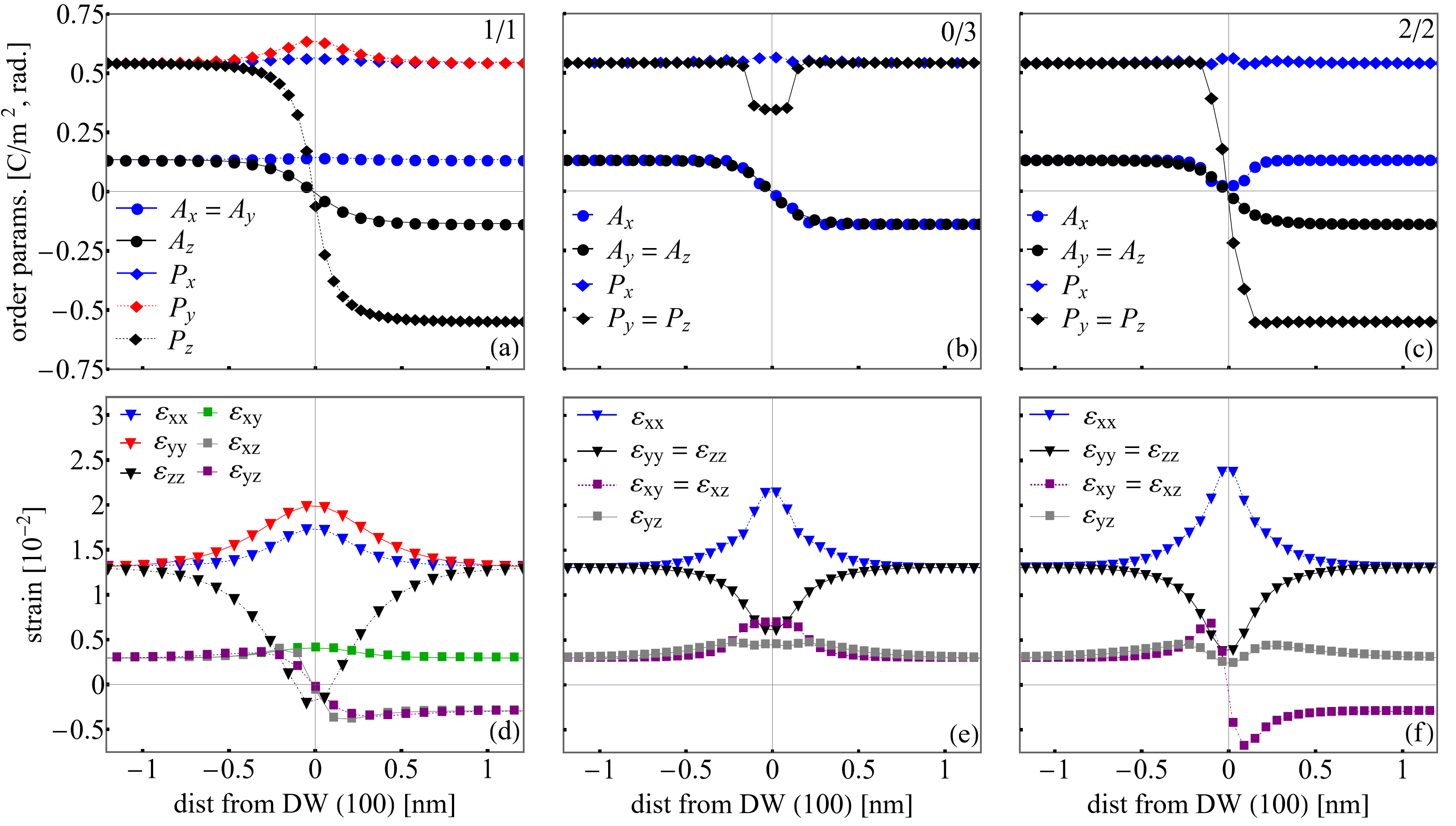}
\caption{\label{fig:1-1_2-2_0-3_profiles} $\mathbf{P}$-$\mathbf{A}$ profiles in arclengths perpendicular to the (100)-oriented DW plane for the (a) 1/1 ($71^\circ$), (b) 0/3 ($180^\circ$ in $\mathbf{A}$), and (c) 2/2 ($109^\circ$) type boundaries. Below in (d), (e),and (f) are the spontaneous strain fields for the normal and shear components along the same arclength. Far from the DW, the solutions converge to the values ($P_s, A_s, \varepsilon_s, \varepsilon_n$), of the ground state.}
\end{figure*}

Next, we present $\mathbf{P}$-$\mathbf{A}$ profiles of three higher energy (100)-oriented domain walls (1/1, 0/3, and 2/2) in Fig~\ref{fig:1-1_2-2_0-3_profiles} (a), (b) and (c) respectively.
These three calculations correspond to those using our best estimates of the gradient coefficients $\{G_{ij},H_{ij}\}$ in Table \ref{tab:gradCoeff}.
In all three cases, we find the presence of a small changes in the non-switching components of the order parameters shown in circles for $A_k$ and diamonds for $P_k$.
For example, in the $71^\circ$ 1/1 shown in DW Fig~\ref{fig:1-1_2-2_0-3_profiles} (a), $P_y$ (in red) which does not change sign, grows at the DW by about 15$\%$.
This is in contrast to the $P_x$ component (in blue) which only grows by 2.5$\%$ demonstrating the influence of the weak built-in field which reduces the magnitude of this component to keep this wall neutral.
Similar changes on the order of about 10$\%$ are also seen in $A_x = A_y$ components shown in blue.
This DW-induced change in $\textbf{P}$ seems to be the largest in the 0/3-type DW shown in (b).
Due to the influence of built in electric fields from the solution of the Poisson equation (and our best estimates of the anisotropic gradient coefficients), the value of $P_x$ component grows by about 5$\%$ whereas the $P_y = P_z$ components \emph{diminish} by almost -35$\%$ (shown in black).
%
%
Again, we also find changes in the non-switching components in the $109^\circ$ 2/2 wall, with $P_x$ (blue diamonds) growing by about 2$\%$; by contrast and $A_x$ decreases by $-6.4^\circ$ (blue circles).
%
%

%
In panels (d), (e), and (f) of Fig.~\ref{fig:1-1_2-2_0-3_profiles}, we depict the corresponding spontaneous strain profiles corresponding to the cases in panels (a), (b), and (c) respectively.
Importantly, far from the DW plane, the spontaneous values of the normal (triangles) and shear (squares) components of the strain converge to their respective values of the single domain state.
However, the strained state of the DW causes various components of $\varepsilon_{ij}$ to grow or depress by large percentages to accommodate the electro- and rotostrictive coupling intrinsic in this structure.
In the case of the 1/1 DW in (d), the value of the $\varepsilon_{zz}$ (in black) shrinks until eventually changing sign (smoothly) at the domain boundary.
For the 2/2 DW, there is a large tensile strain in $\varepsilon_{xx}$ (in blue) growing by about a factor of three across the wall.
Also presented in Table \ref{tab:FEdw1} are the DW thicknesses associated to the corresponding order parameters, which differ between $\mathbf{P}$ and $\mathbf{A}$.
We should note that the thicknesses of the DW corresponding to $\mathbf{P}$ and $\mathbf{A}$ differ.
This arises because our resulting fit parameters are anisotropic (i.e. $H_{11} \ll H_{44}$) and also the presence of growth/decrease in non-switching components of $\mathbf{P}$ and $\mathbf{A}$ due to the roto- and electrostrictive coupling.
Nevertheless, as seen in the table, the domain walls are quite thin ($2 t_k \approx 0.05 - 0.5$ nm) which agrees quite-well with the available literature on BFO suggesting atomistically thin DWs \cite{Dieguez2013, Hlinka2017, Korbel2020, Borisevich2010}.
The smaller value of the DW thickness in the $\mathbf{A}$ (as compared to $\mathbf{P}$) also shows good qualitative agreement with measurements from experiments using Z-contrast scanning transmission electron microscopy\cite{Borisevich2010}.



%
We extend this type of analysis iteratively for the possible DWs listed in Table \ref{tab:FEdw1} so that we can converge our set of coefficients yielding reasonable $F_\mathrm{DW}$ values comparable to DFT; importantly, capturing the energy hierarchy\cite{Dieguez2013, Korbel2020, Xue2014} predicted for the collection of walls.
Our best estimates of the gradient coefficients found through our fitting procedure are presented in Table~\ref{tab:gradCoeff}.
We find that $H_{11} \ll -H_{12} < H_{44}$ in agreement with similar studies on BFO\cite{Xue2014, Xue2021}. 
This is an important relationship that results from harmonic models of antiferrodistortive cubic perovskite materials which has been connected to an asymmetry in the phonon bands at the R point\cite{Gesi1972, Stirling1972, Cao1990}.
Another result from our fits is that the energy hierarchy yields $F_\mathrm{DW}(109^\circ) < F_\mathrm{DW}(180^\circ) < F_\mathrm{DW}(71^\circ)$  for the lowest energy walls\cite{Lubk2009, Dieguez2013, Xue2014, Xue2021}.

\begin{table}[h!]
\caption{\label{tab:gradCoeff}
Best estimates of the six independent lowest-order Lifshitz invariant coefficients $G_{ij}$ and $H_{ij}$ found through our fitting procedure. Units are given in $[10^{-9}\mathrm{J}\cdot\mathrm{m}^3\cdot \mathrm{C}^{-2}]$ and $10^{-9}\mathrm{J}\cdot\mathrm{m}^3\cdot\mathrm{deg}^{-2}]$ respectively.
}
\begin{ruledtabular}
\begin{tabular}{c c c c c c c c c c c c c }
 & $H_{11}$ & & $H_{12}$ & & $H_{44}$ & & $G_{11}$ & &  $G_{12}$ & & $G_{44}$    \\ 
  & $0.005$ & & $-1.0$ & & $4.0$  & & $28.0$ & & $-15.0$ & & $0.5$    \\ 
\end{tabular}
\end{ruledtabular}
\end{table}

\subsection{Antiferromagnetic energy terms}\label{sec:magE}

Now we turn to the AFM order present in BFO. 
To encapsulate the magnetic behavior of single crystalline BFO, we propose a continuum-approximation to the magnetic free energy density. 
We consider the total free energy density of the magnetic subsystem ($f_\mathrm{mag}$) to be a sum of the terms responsible for the nominally collinear AFM sublattices ($f_\mathrm{sp}$) and those producing the noncollinearity (canted magnetism) by coupling to the structural order ($f_\mathrm{MP}$).
We first consider the magnetic energy due to the spin subsystem that is not coupled to the structural order,
\begin{equation}\label{eqn:AFM}
\begin{aligned}
    f_\mathrm{sp} =& D_e \left(\mathbf{L}^2 - \mathbf{m}^2\right) \\
    +& A_e \left[\left(\nabla L_x\right)^2 + \left(\nabla L_y\right)^2 + \left(\nabla L_z\right)^2\right] \\
    +&\sum\limits_{\eta = 1}^2 K_1^c \left(m_{\eta,x}^2 m_{\eta,y}^2 m_{\eta,z}^2\right), \\
\end{aligned}
\end{equation}
where $D_e < 0$ controls the strength of the short-range superexchange energy which favors the spins to have collinear AFM ordering \cite{Rezende2019}.
At our coarse-grained level of theory, we only consider the first nearest-neighbor exchange coupling which has been calculated from first-principles methods\cite{Xu2019} to be approximately 6 meV/f.u. corresponding to $D_e = -23.4 \,\mathrm{meV}/\mathrm{nm}^3$ in our simulations.
The second term describes the AFM non-local exchange stiffness proposed in Ref. [\citen{Agbelele2017}] with $A_e = 18.7$ meV/nm (or $3\times10^{-7}$ ergs/cm).
The third term corresponds to a weak single-ion anisotropy \cite{Weingart2012} with $K_1^c = 2.2 \times 10^{-3} \,\mu$eV/$\mathrm{nm}^{-3}$; this term reflects the cubic symmetry of the lattice and breaks the continuous degeneracy of the magnetic easy-plane into a six-fold symmetry.
The remaining terms are due to the magnetostructual coupling,
\begin{equation}\label{eqn:FmagMP}
\begin{aligned}
    f_\mathrm{MP} &= f_\mathrm{DMI}(\mathbf{A}) + f_\mathrm{easy}(\mathbf{P}) + f_\mathrm{anis}(\mathbf{A}),
\end{aligned}
\end{equation}
where
\begin{equation}\label{eqn:dmi}
\begin{aligned}
    f_\mathrm{DMI} = D_0 \textbf{A} \cdot \left( \textbf{L} \times \textbf{m}\right),
\end{aligned}
\end{equation}
is due to the antisymmetric DMI which acts to break the collinearity by competing energetically with the first term of Eq.~(\ref{eqn:AFM}).
It should be emphasized here that the local oxygen octahedral environments of adjacent Fe atoms underpins the DMI \emph{vector} \cite{Ederer2005, Fennie2008, deSousa2009, Rahmedov2012, Meyer2022}. 
Therefore, the $\mathbf{A}$ order parameter enables the DMI coupling.
Ref. [\citen{Dixit2015}] provides an estimate of the DMI energy corresponding to $304$ $\mu$eV/f.u. 
It should be mentioned that the weak canting between $\mathbf{m}_1$ and $\mathbf{m}_2$ arises from a competition between $D_e$ and $D_0$ and that different estimates of their values can provide the same degree of canting of the sublattices provided they have the same ratio $D_e/D_0$.
We come back to this in the next section.

BFO is an easy-plane antiferromagnet\cite{Dixit2015}, in which the magnetic sublattices lie in a plane defined by the direction of $\mathbf{P}$.
We include the magnetocrystalline anisotropy term\cite{Rezende2019} requisite for easy-plane AFMs as,
\begin{equation}\label{eqn:easy}
\begin{aligned}
    f_\mathrm{easy} = \sum\limits_{\eta = 1}^2 K_1 \left(\textbf{m}_\eta \cdot \hat{\textbf{P}}\right)^2
\end{aligned}
\end{equation}
with the usual definition of $K_1 > 0$ enforcing the easy-plane condition for $\mathbf{m}_\eta$ with $\eta = 1,2$.
Using DFT methods, Dixit and co-workers \cite{Dixit2015}, determined that the relative energy difference between aligning the magnetic sublattices along $\mathbf{P}$ or in the plane normal to $\mathbf{P}$ is $-2.0$ meV/f.u.
Therefore, we choose $K_1 = 31.25 \mathrm{meV}/\mathrm{nm}^3$ for our simulations.
We further couple the magnetic energy surface to the structural order by allowing the weak single-ion anisotropy to also depend on the antiphase tilts $\mathbf{A}$ \cite{Weingart2012} through,
\begin{equation}\label{eqn:anis}
\begin{aligned}
    f_\mathrm{anis} &=  \sum\limits_{\eta = 1}^2 a|\textbf{A}|^2\left(m_{\eta,x}^2 m_{\eta,y}^2 m_{\eta,z}^2\right)
\end{aligned}
\end{equation}
which is in addition to the term in Eq.~(\ref{eqn:AFM}).
The choice of $K_1^c > 0$ and $0 < a |\textbf{A}|^2 < K_1^c$ corresponds to a small energy barrier between the 6-fold possible orientations of the weak magnetization $\mathbf{m}$ thus breaking the continuous degeneracy in the easy-plane. 
These coefficients can be obtained from DFT calculations as shown in Refs [\citen{Dixit2015}] and [\citen{Weingart2012}].
Therefore, we choose our coefficients (see Table \ref{tab:spincoeff}) such that the relative energy density barrier for the six-fold symmetry is 0.01 meV/$\mathrm{nm}^3$ which is a reasonable approximation based on the aforementioned works.
We find no influence of this choice of coupling constant on the results presented in this manuscript.
%
%
The coefficients for $f_\mathrm{sp}$ and $f_\mathrm{MP}$ are listed in Table \ref{tab:spincoeff}.
\begin{table}[h!]
\caption{\label{tab:spincoeff}%
Spin free energy density materials coefficients used in this work.
}
\begin{ruledtabular}
\begin{tabular}{c c c c c c c c c c c c c    }
 & & & $A_e$ & & 18.7 & & $[\mathrm{meV} \mathrm{nm}^{-3}]$ & & Ref. [\citen{Agbelele2017}] & &  &     \\ 
 & & & $D_e$ & & -23.4 & & $[\mathrm{meV} \mathrm{nm}^{-3}]$ & & Ref. [\citen{Xu2019}] & &  &     \\ 
 & & & $D_0$ & & 0.0046 & & $[\mathrm{meV} \mathrm{deg}^{-1} \mathrm{nm}^{-3}]$  & & this work & &  &     \\ 
 & & & $K_1$ & & 31.25 & & $[\mathrm{meV} \mathrm{nm}^{-3}]$ & & Ref. [\citen{Dixit2015}] & &  &     \\ 
 & & & $K_1^c$ & & 0.0022 & & $[\mathrm{meV} \mathrm{nm}^{-3}]$ & &  & &  &     \\ 
 & & & $a$ & & 0.00015 & & $[\mathrm{meV} \mathrm{deg}^{-1} \mathrm{nm}^{-3}]$ & &  & &  &     \\ 
\end{tabular}
\end{ruledtabular}
\end{table}
We should note that a long-period ($\lambda \approx 64$ nm) cycloidal rotation of the weak magnetization is often observed in BFO samples\cite{Burns2020, Agbelele2017, Xu2021}.
It is possible to eliminate the cycloidal order by doping \cite{Sosnowska2002}, epitaxial strain \cite{Bai2005, Sando2013, Burns2020}, applied electric fields \cite{deSousa2013}, or by some processing techniques (i.e. via a critical film thickness) \cite{Burns2020} during synthesis.
The spin-cycloid could be incorporated into our model by including coupling terms associated with a proposed spin-current mechanism \cite{Agbelele2017, Popkov2016, Xu2022}. However, in order to provide the simplest model of the ME multiferroic effects, we have neglected them in this work.
\subsection{Micromagnetics and homogeneous spin ground states}\label{sec:magHo}
In order to find the spin ground states in the presence of an arbitrary structural fields, we consider the Landau-Lifshitz-Bloch (LLB) equation \cite{Garanin2004} that governs the sublattices $\mathbf{m}_\eta$, 
\begin{equation}\label{eqn:LLG_LLB}
\begin{aligned}
    \frac{d \mathbf{m}_\eta}{dt}&=-\frac{\gamma}{1+\alpha^2}\left(\mathbf{m}_\eta\times\mathbf{H}_\eta\right) \\
    &-\frac{\gamma\alpha}{1+\alpha^2}\mathbf{m}_\eta\times\left(\mathbf{m}_\eta\times\mathbf{H}_\eta\right) \\
    &+ \frac{\gamma\tilde\alpha_\parallel} {(1+\alpha^2)} m_\eta^2 \left[m_\eta^2 - 1\right]\mathbf{m}_\eta. \\
\end{aligned}
\end{equation}
where $\alpha$ is the phenomenological Gilbert damping parameter and $\gamma$ is the electronic gyromagnetic coefficient equal to $2.2101\times10^5$~rad.~m~A${}^{-1}$~s${}^{-1}$. 
The effective fields are defined as $\mathbf{H}_\eta = - \mu_0^{-1} M_s^{-1} \delta f / \delta \mathbf{m}_\eta$ with $\mu_0$ the permeability of vacuum. The saturation magnetization density of the BFO sublattices is $M_s = 4.0$ $\mu$B/Fe\cite{Weingart2012,Xu2019,Xu2022}.
The third term arises from the LLB approximation in the zero temperature limit where $\tilde\alpha_\parallel$ is a damping along the longitudinal direction of $\mathbf{m}_\eta$.
We implement the LLB equation as a numerical resource in order to provide a restoring force and bind the quantities $\mathbf{m}_\eta$ to the unit sphere $(|\mathbf{m}_\eta| = 1)$.
In this context, we consider our spin subsystem to be at $T = 0$ K in results presented throughout in this paper.
In the Appendix, we provide a short derivation of the LLB torque in the zero temperature limit.
We set $\tilde\alpha_\parallel = 10^3$ in all results in this work to satisfy the constraint on $\mathbf{m}_\eta$.

\begin{figure}[h!]\centering
\hspace*{-10pt}\includegraphics[scale=0.36]{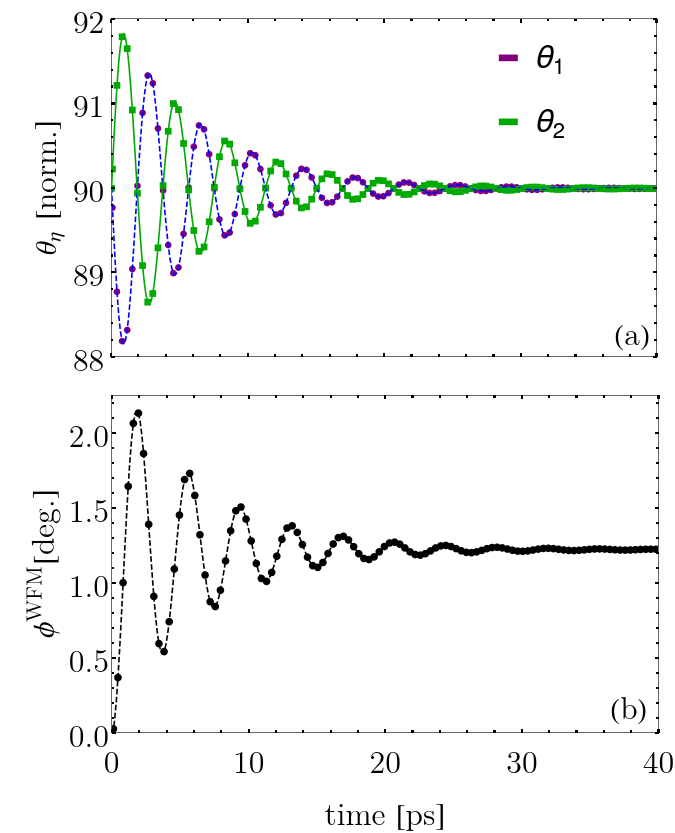}
\caption{\label{fig:1ringdown} (a) easy plane angles $\theta_\eta$ and (b) canted moment angle $\phi^\mathrm{WFM}$ during the magnetic ringdown of Eq.~(\ref{eqn:LLG_LLB}) with $\alpha = 0.05$ and $\mathbf{P}\uparrow\uparrow\mathbf{A}$ aligned along the [111] direction. The longitudinal damping in the LLB equation enforces the normalization $|\mathbf{m}_1| = |\mathbf{m}_2| = 1$ at all time steps in the evolution.}
\end{figure}

To look for homogeneous spin ground states, we consider $\alpha = 0.05$ and evolve Eq.~\ref{eqn:LLG_LLB} (utilizing the numerical approach described in Sec. \ref{sec:num}) until the relative change in the total energy computed from the summation of Eq.~(\ref{eqn:AFM}) and Eq.~(\ref{eqn:FmagMP}) between adjacent time steps is $\Delta F < 10^{-8}$ eV/$\mu$s.
Also, we stress that the influence of $\tilde\alpha_\parallel$ is negligible in all results presented in this work provided that its unitless value is around $10^3$ or above. 
To verify that our ground states predict the magnetic ordering consistent with the literature of BFO, we define two angular variables $\phi^\mathrm{WFM} = \cos^{-1}{\left(\mathbf{m}_1 \cdot \mathbf{m}_2\right)}$ and $\theta_\eta = \cos^{-1}{\left(\mathbf{m}_\eta \cdot \hat{\mathbf{P}}\right)}$.
The former tracks the degree of canting between the sublattices and the latter tracks the orientation of the magnetization with respect to $\hat{\mathbf{P}} = \mathbf{P}/P_s$, the magnetic easy-plane normal.
As an example, we first set $\mathbf{P}\uparrow\uparrow\mathbf{A}$ along the [111] direction to be static.
The time evolution (ringdown) of Eq.~(\ref{eqn:LLG_LLB}) is highlighted in Fig.~\ref{fig:1ringdown}(a) for $\theta_\eta$ showing that the sublattices have relaxed into the easy plane defined by $\hat{\mathbf{P}}$ with $\theta_1 = \theta_2 = 90.0^\circ$
In (b) the time dependence of the canting angle $\phi^\mathrm{WFM}$ during the relaxation is shown. 
At the conclusion of the ringdown, $\phi^\mathrm{WFM}$ reaches a value of $\approx 1.22^\circ)$.
%
%
This demonstrates that the angular quantities $\{\theta_\eta, \phi^\mathrm{WFM}\}$ detail an orthogonal system of the $\{ \textbf{P}, \textbf{m}, \textbf{L}  \}$ vectors as often discussed in the literature \cite{Heron2014}.

\begin{table}[h!]
\caption{\label{tab:magGs}%
Six-fold symmetric magnetic ground states for each ($\mathbf{P}\uparrow\uparrow\mathbf{A}$) domain orientation. Note that these listed directions are not corrected for the DMI interaction and therefore $\mathbf{m}_1 \neq - \mathbf{m}_2$ (hence $\simeq$). All dot products yield an orthogonal system for $\{ \textbf{P}, \textbf{m}, \textbf{L}  \}$. Full reversal of $\mathbf{A}$ changes the sign on $\mathbf{m}$ but not $\mathbf{L}$. The small corrections, due to DMI, are on the order of the canting angle $\phi^\mathrm{WFM}$ ($\approx 1.22^\circ$).
}
\begin{ruledtabular}
\begin{tabular}{c c c c c c c c c c c c c c c c c}
 ($\mathbf{P}\hspace{-1pt}\uparrow\uparrow\hspace{-1pt}\mathbf{A}$): & $\hspace{-5pt}[111]$ & & $\hspace{-5pt}[\bar{1}11]$& & $\hspace{-5pt}[1\bar{1}\bar{1}]$& & $\hspace{-5pt}[\bar{1}\bar{1}\bar{1}]$ &  & $\hspace{-5pt}[1\bar{1}1]$ & & $\hspace{-5pt}[11\bar{1}]$&  & $\hspace{-5pt}[\bar{1}\bar{1}1]$& & $\hspace{-5pt}[\bar{1}1\bar{1}]$& \\ 
\colrule
& & & & & & & & & & & & & & & & \\
$\mathbf{L} \simeq$ & $\hspace{-5pt}[\bar{1}10]$  & & $\hspace{-5pt}[101]$ & & $\hspace{-5pt}[101]$ & & $\hspace{-5pt}[\bar{1}10]$ & & $\hspace{-5pt}[110]$ & & $\hspace{-5pt}[\bar{1}10]$ & & $\hspace{-5pt}[\bar{1}10]$ & & $\hspace{-5pt}[110]$ &\\ 
& $\hspace{-5pt}[\bar{1}01]$ & & $\hspace{-5pt}[110]$ & & $\hspace{-5pt}[110]$ & & $\hspace{-5pt}[\bar{1}01]$ & & $\hspace{-5pt}[011]$ & & $\hspace{-5pt}[011]$ & & $\hspace{-5pt}[011]$ & & $\hspace{-5pt}[011]$ &\\ 
& $\hspace{-5pt}[0\bar{1}1]$ & & $\hspace{-5pt}[01\bar{1}]$& & $\hspace{-5pt}[01\bar{1}]$& & $\hspace{-5pt}[0\bar{1}1]$ & & $\hspace{-5pt}[\bar{1}01]$ & & $\hspace{-5pt}[101]$ & & $\hspace{-5pt}[101]$ & & $\hspace{-5pt}[\bar{1}01]$&\\ 
&  $\hspace{-5pt}[1\bar{1}0]$ & & $\hspace{-5pt}[\bar{1}0\bar{1}]$ & & $\hspace{-5pt}[\bar{1}0\bar{1}]$ & & $\hspace{-5pt}[1\bar{1}0]$  & & $\hspace{-5pt}[\bar{1}\bar{1}0]$ & & $\hspace{-5pt}[1\bar{1}0]$ & & $\hspace{-5pt}[1\bar{1}0]$ & & $\hspace{-5pt}[0\bar{1}\bar{1}]$ &\\ 
&  $\hspace{-5pt}[10\bar{1}]$ & & $\hspace{-5pt}[\bar{1}\bar{1}0]$ & & $\hspace{-5pt}[\bar{1}\bar{1}0]$ & & $\hspace{-5pt}[10\bar{1}]$ & & $\hspace{-5pt}[0\bar{1}\bar{1}]$ & & $\hspace{-5pt}[0\bar{1}\bar{1}]$ & & $\hspace{-5pt}[0\bar{1}\bar{1}]$ & & $\hspace{-5pt}[0\bar{1}\bar{1}]$ &\\ 
&  $\hspace{-5pt}[01\bar{1}]$ & & $\hspace{-5pt}[0\bar{1}1]$ & & $\hspace{-5pt}[0\bar{1}1]$ & &  $\hspace{-5pt}[01\bar{1}]$ & & $\hspace{-5pt}[10\bar{1}]$ & & $\hspace{-5pt}[\bar{1}0\bar{1}]$ & & $\hspace{-5pt}[\bar{1}0\bar{1}]$ & & $\hspace{-5pt}[10\bar{1}]$&\\ 
\colrule
& & & & & & & & & & & & & & & & \\
$\mathbf{m} \simeq$ & $\hspace{-5pt}[\bar{1}\bar{1}2]$  & & $\hspace{-5pt}[12\bar{1}]$&  &$\hspace{-5pt}[\bar{1}\bar{2}1]$ & & $\hspace{-5pt}[11\bar{2}]$ & & $\hspace{-5pt}[\bar{1}12]$ & & $\hspace{-5pt}[112]$ & & $\hspace{-5pt}[\bar{1}\bar{1}\bar{2}]$ & & $\hspace{-5pt}[1\bar{1}\bar{2}]$ \\
 & $\hspace{-5pt}[1\bar{2}1]$ & & $\hspace{-5pt}[\bar{1}1\bar{2}]$ & & $\hspace{-5pt}[1\bar{1}2]$ & & $\hspace{-5pt}[\bar{1}2\bar{1}]$ & & $\hspace{-5pt}[\bar{2}\bar{1}1]$ & & $\hspace{-5pt}[2\bar{1}1]$& & $\hspace{-5pt}[\bar{2}1\bar{1}]$ & & $\hspace{-5pt}[21\bar{1}]$\\
 & $\hspace{-5pt}[2\bar{1}\bar{1}]$ & & $\hspace{-5pt}[\bar{2}\bar{1}\bar{1}]$ &  & $\hspace{-5pt}[211]$ & & $\hspace{-5pt}[\bar{2}11]$ & & $\hspace{-5pt}[\bar{1}\bar{2}\bar{1}]$& & $\hspace{-5pt}[1\bar{2}\bar{1}]$ & & $\hspace{-5pt}[\bar{1}21]$ & & $\hspace{-5pt}[121]$\\
 & $\hspace{-5pt}[11\bar{2}]$ & & $\hspace{-5pt}[\bar{1}\bar{2}1]$ &  & $\hspace{-5pt}[12\bar{1}]$ & & $\hspace{-5pt}[\bar{1}\bar{1}2]$ & & $\hspace{-5pt}[1\bar{1}\bar{2}]$ & & $\hspace{-5pt}[\bar{1}\bar{1}\bar{2}]$ & & $\hspace{-5pt}[112]$ & & $\hspace{-5pt}[\bar{1}12]$\\
 & $\hspace{-5pt}[\bar{1}2\bar{1}]$ & & $\hspace{-5pt}[1\bar{1}2]$ & & $\hspace{-5pt}[\bar{1}1\bar{2}]$ & & $\hspace{-5pt}[1\bar{2}1]$ & & $\hspace{-5pt}[21\bar{1}]$ & & $\hspace{-5pt}[\bar{2}1\bar{1}]$ & & $\hspace{-5pt}[2\bar{1}1]$ & & $\hspace{-5pt}[\bar{2}\bar{1}1]$\\
 & $\hspace{-5pt}[\bar{2}11]$ & & $\hspace{-5pt}[211]$ &  & $\hspace{-5pt}[\bar{2}\bar{1}\bar{1}]$ & & $\hspace{-5pt}[2\bar{1}\bar{1}]$ & & $\hspace{-5pt}[121]$ & & $\hspace{-5pt}[\bar{1}21]$ & & $\hspace{-5pt}[1\bar{2}\bar{1}]$ &  & $\hspace{-5pt}[\bar{1}\bar{2}\bar{1}]$\\
\end{tabular}
\end{ruledtabular}
\end{table}

\begin{figure}[h!]
\includegraphics[scale=0.31]{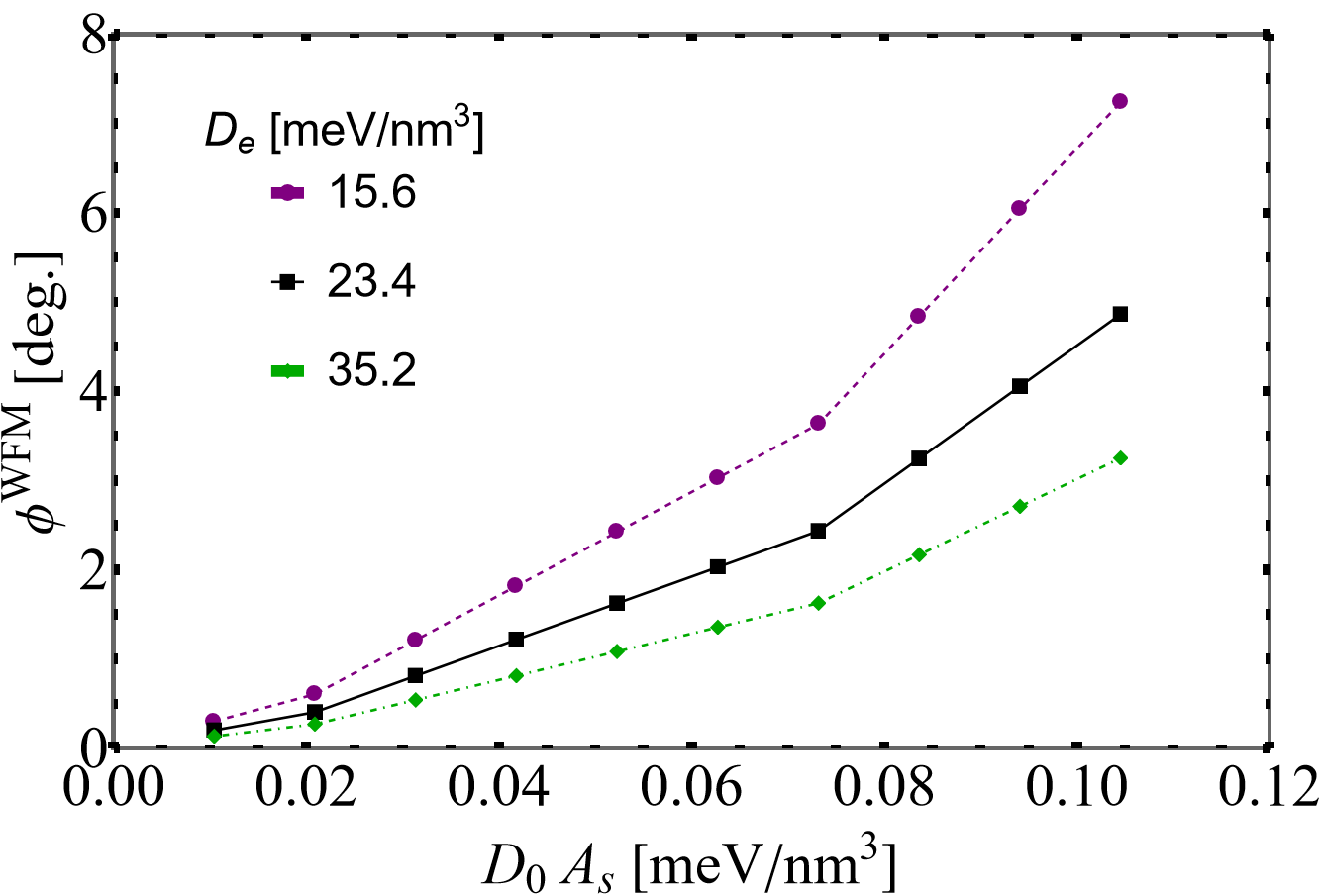}
\caption{\label{fig:2phi} Dependence of canting angle $\phi^\mathrm{WFM}$ on the ground state DMI free energy density $(D_0 A_s)$ for different choices of the AFM superexchange parameter $D_e$.}
\end{figure}
As a further benchmark, we probe the influence of the ratio $D_e/D_0$ on the values of $\phi^\mathrm{WFM}$.
This test, shown in Fig.~\ref{fig:2phi} highlights the energetic competition between the AFM superexchange and the sublattice DMI.
%
%
From Ref. [\citen{Xu2019}] we have $D_e = 23.4$ meV/nm${}^3$ and our analysis demonstrates $\phi^\mathrm{WFM} = 1.22^\circ$ provided $D_0 A_s = 0.036$ meV/nm${}^3$.
We thus have the weak moment $M_s |\mathbf{m}| = 0.03$ $\mu$B/Fe, which agrees well with the available literature \cite{Wojdel2010, Tokunaga2010, Dixit2015, Xu2019}.
By setting $\mathbf{P}\uparrow\uparrow\mathbf{A}$ along the eight polar directions possible in BFO, we can find six magnetic states for each of them.
The corresponding 48 multiferroic domains are listed in Table \ref{tab:magGs}.
For all $\mathbf{m}$ orientations calculated, the canted angle is precisely $\phi^\mathrm{WFM} = 1.22^\circ$.
%
%
Additionally, when $\mathbf{A}$ is reversed fully ($\mathbf{P}\uparrow\downarrow\mathbf{A}$), which is an acceptable ground state in our potential, the sign of $\mathbf{m}$ will change but not the sign of the Néel vector $\mathbf{L}$. Hence, we have a total of 96 possible domain variants. Due to the DMI these quantities listed in this table are canted slightly from their listed values (hence the use of $\simeq$ symbol).
\begin{figure*}\centering
\hspace*{-10pt}\includegraphics[scale=0.36]{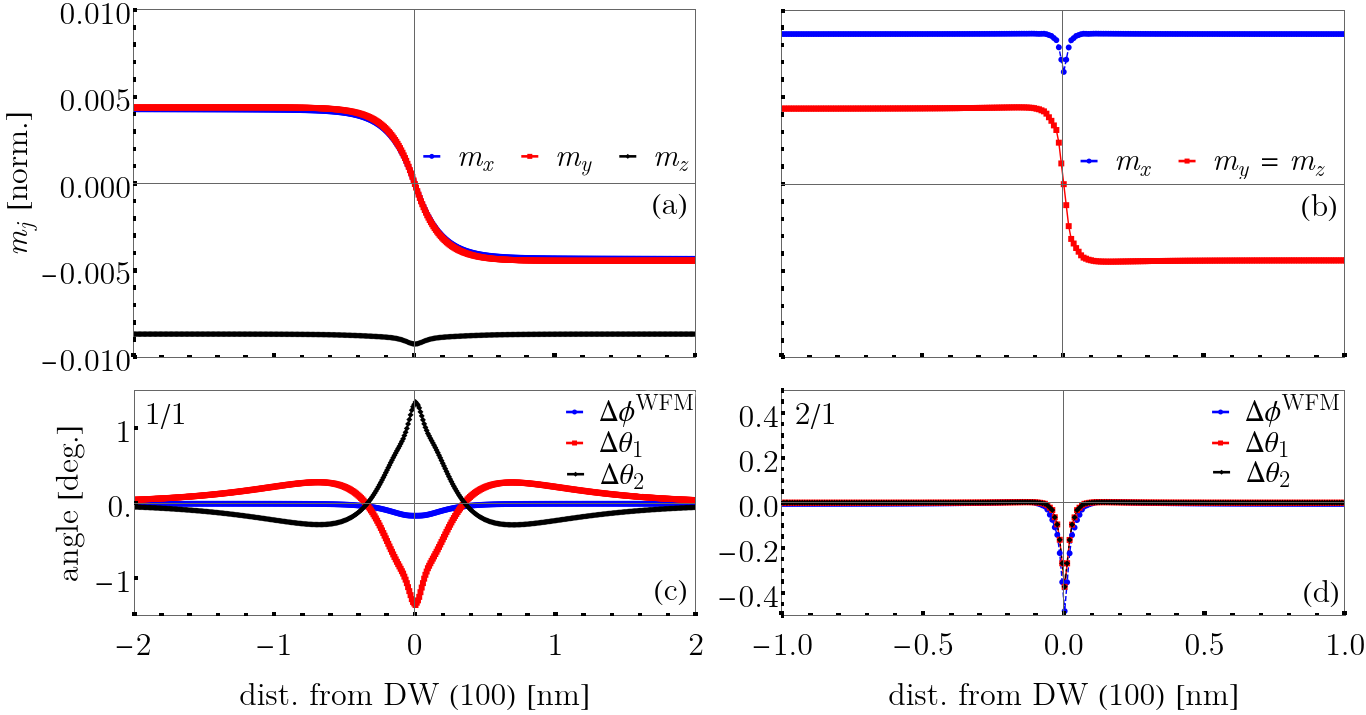}
\caption{\label{fig:mag_walls} Net magnetization $\mathbf{m}$ textures presented in normalized units across the (a) 1/1, and (b) 2/1 DWs of (100)-orientation. Both of these sequences of DWs produce $71^\circ$ rotations of $\mathbf{m}$. Angular deviations from the ground state values of $\phi^\mathrm{WFM}$, $\theta_1$, and $\theta_2$ for 1/1 (c) indicate a much longer range coupling of the spin across the ME boundary than in the 2/1 case in (d).}
\end{figure*}

\subsection{Antiferromagnetic domain walls}\label{sec:magDW}

%
%
Using low-energy electron microscopy (X-PEEM), AFM domain boundary contrast can be visualized\cite{Moubah2012} within a single ferroelectric domain.
To better understand the capabilities of our modeling effort, we attempt to stabilize an AFM DW (i.e., one with switched $\mathbf{L}$) corresponding to the above experimental observations.
We set $\mathbf{P}\uparrow\uparrow \mathbf{A}$ along $[11\bar{1}]$ to be homogeneous (and fixed in time) within the computational box. 
Then, a sin(x) profile is chosen for the sublattices $\mathbf{m}_\eta$ corresponding to two possible N\'{e}el orientations of Table \ref{tab:magGs} for a (100)-oriented domain boundary with homogeneous $\mathbf{P}$.
After relaxation Eq.~(\ref{eqn:LLG_LLB}) with large Gilbert damping $\alpha = 0.8$, we find that the AFM wall is not stable and the system evolves to a homogeneous state with $\mathbf{L}$ corresponding to one of the six possible orientations allowed in the domain. 
If the non-local exchange interaction governed by $A_e$\cite{Agbelele2017} is reduced by a factor of ten, then we find that the solution corresponds to AFM domain walls with a $120^\circ$ rotation of $\mathbf{L}$, i.e $\mathbf{L}^\mathrm{I} = [011]$ and $\mathbf{L}^\mathrm{II} = [1\bar{1}0]$.
We estimate that the corresponding DW in $\mathbf{L}$ has a characteristic width of 20 nm and a corresponding DW energy of $7.55$ mJ/$\mathrm{m}^2$ using Eq.~(\ref{eqn:enDW}).

Let us now consider how the structural DWs affect the net magnetization.
The modulation of $\mathbf{P}$ and $\mathbf{A}$ across the domain boundary drastically alters the magnetostructural coupling energy surface due to Eq.~(\ref{eqn:FmagMP}) causing the AFM order to choose preferential orientations associated with those calculated in Table \ref{tab:magGs}.
Careful inspection of Table \ref{tab:magGs} suggests that only certain low energy magnetic DWs (i.e., those minimizing the gradient of $\mathbf{L}$) should be observed for the different FE domain walls listed in Table \ref{tab:FEdw1}.
Using our previously established notation for adjacent DW states, the lowest energy FE DW (2/1) corresponding to a $\mathbf{P}^\mathrm{I}/\mathbf{A}^\mathrm{I} = [\bar{1}11]/[\bar{1}11]$ to $\mathbf{P}^\mathrm{II}/\mathbf{A}^\mathrm{II} = [\bar{1}\bar{1}\bar{1}]/[111]$ change will only allow $\mathbf{m}^\mathrm{I} = [211]$ or $\mathbf{m}^\mathrm{I} = [\bar{2}\bar{1}\bar{1}]$ and $\mathbf{m}^\mathrm{II} = [2\bar{1}\bar{1}]$ or $\mathbf{m}^\mathrm{II} = [\bar{2}11]$ respectively with no changes to the Néel vector $\mathbf{L}$.
This coincides with a $71^\circ$ rotation of $\mathbf{m}$ consistent with a $71^\circ$ change of the oxygen octahedral tilt field $\mathbf{A}$ albeit having a $109^\circ$ $\mathbf{P}$ switch.
%

%
%

%
To calculate the magnetic textures numerically, we fix in time the FE order parameters $\mathbf{P}$-$\mathbf{A}$ corresponding to a specific DW in Sec. \ref{sec:DWparam}.
We choose the 1/1 (100) and 2/1 (100) structural walls as they are most commonly observed in experiment.
Again, we use a large Gilbert damping $\alpha = 0.8$ and look for the ground states utilizing Eq.~(\ref{eqn:LLG_LLB}).
%
%
In Fig.~\ref{fig:mag_walls}, we display the weak $\mathbf{m}$ moment as a function of the distance to the DW plane for the 1/1 (a) and 2/1 (b) walls after relaxation.
In both cases, the $\mathbf{m}$ rotates by $71^\circ$ - $[11\bar{2}]$ to $[\bar{1}\bar{1}\bar{2}]$ in (a) and $[211]$ to $[2\bar{1}\bar{1}]$ in (b) - with a sharp interface region.  
This is expected as the DMI term is driven by the $\mathbf{A}$ vector forcing $\mathbf{m}$ to also change by $71^\circ$.
The large value of $A_e$ causes the N\'{e}el vector to be nearly constant across the DW corresponding to $[1\bar{1}0]$ in (a) and $[0\bar{1}1]$ in (b) as it satisfies both conditions of the ground state in adjacent domains. 
Fitting the switched components of $\mathbf{m}$ to the aforementioned tanh(x) profile from Sec. \ref{sec:DWparam} yields $2 t_m = 0.5$ nm. 
We can calculate a thickness of $2 t_m = 0.06$ nm in the 2/1 (100) case demonstrating a nearly atomistically thin DW in the magnetic texture.
A comparison to Table \ref{tab:FEdw1} shows that we have an equality of $t_m \approx t_A$ in both 1/1 (100) and 2/1 (100) walls.
%
%

%
The component of $\mathbf{m}$ that does not switch, black in (a) and blue in (b), changes by about $\approx +6\%$ and $-20\%$ respectively across the DW region indicating rotational components of $\mathbf{m}$.
This leads to a deviations of the angular quantities $\{\phi^\mathrm{WFM},\theta_1,\theta_2\}$ from their ground state values.
We plot these quantities in panels (c) and (d) of Fig.~\ref{fig:mag_walls}.
We see that, in the 1/1 (100) case in (c), the sublattices cant slightly ($\approx \pm 1^\circ$) out of the easy plane to facilitate this magnetic reversal.
The weak magnetization canting angle $\phi^\mathrm{WFM}$ (shown in blue) also reduces its magnitude by about $0.25^\circ$.
This is different from the behavior of the angular quantities of the 2/1 (100) DW shown in panel (d) which decrease their values by about $0.4^\circ$ in the same fashion indicating canting out of the easy-plane in the same direction for both sublattices resulting in a slight reduction of $\mathbf{m}$.
We stress that these quantities should be meaningful since they are on the order of $\phi^\mathrm{WFM}$ in the ground state and that in the 1/1 (100) case, the modulations extend more than a few unit cells from the DW ($\pm 2$ nm).
By using Eq.~(\ref{eqn:enDW}), we can estimate the energy of the magnetic DW of the 1/1 (100) and 2/1 (100) cases.
For the 1/1 and 2/1 walls, we calculate $F_\mathrm{DW}^\mathrm{mag} = 0.71$ and $0.70$ mJ/$\mathrm{m}^2$ respectively. 
%
%
The energy difference between these two $71^\circ$ $\mathbf{m}$ DWs is quite small despite having a very different profile of $\theta_\eta$ and $\phi^\mathrm{WFM}$.
The variation of $\theta_\eta$ in panel (c) for the 1/1 (100) case causes a large relative increase in the easy-plane anisotropy for both sublattice contributions as compared to (d) for the 2/1 (100) DW.
However, as seen in the panel (d), there is more identifiably sharp structure (i.e., modulations of $\phi^\mathrm{WFM}$ and $\theta_\eta$ occur within $\pm 0.2$ nm of the DW) as $\mathbf{m}$ switches by $71^\circ$.
This leads to an increase in the DMI energy relative to the 1/1 case. 
We have only presented data on these two types of magnetic boundaries in the presence of the $\mathbf{P}$-$\mathbf{A}$ DWs.
Higher energy DWs can also be investigated with our approach, but we leave this for future work.
%
%
%
%

%
\section{Applications: spin waves and magnetoelectric switching}\label{sec:appl}

\subsection{Spin waves through multiferroic domain boundaries}\label{sec:applSpWav}

\begin{figure*}\centering
\hspace*{-10pt}\includegraphics[scale=0.364]{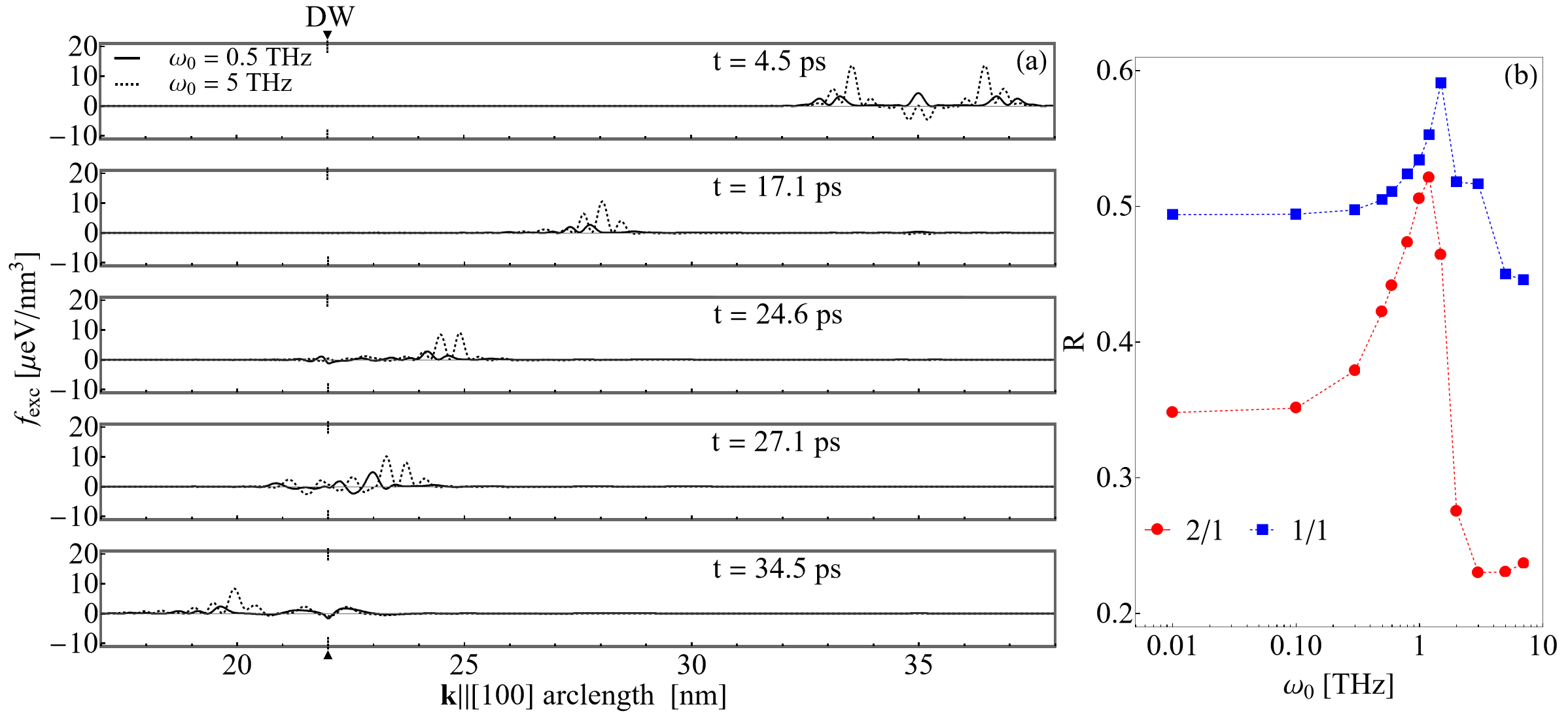}
\caption{\label{fig:spin_pulse} (a) Excess energy density $f_\mathrm{exc}$ due to a spin wave traveling in the $\mathbf{k}||[100]$ direction. The excitation frequency is $\omega_0 = 0.5$ and $5$ THz for the solid and dashed lines respectively. In this simulation, the 2/1 ($109^\circ$) DW located at approximately 22 nm indicated by the arrow. The wavefront reaches the DW at around 27 ps. (b) Calculated spin wave rectification $R$ as a function of $\omega_0$ of the 1/1 (blue) and 2/1 (red) DWs using Eq.~(\ref{eqn:rec}) after time integrating $f_\mathrm{exc}$ at a distance of $\Delta x = 7$ nm left and right from the DW.}
\end{figure*}

The field of spintronics relies on the generation, control, and read-out of traveling packets of spin\cite{Hirohata2020}.
In AFMs, the spin precessional processes can occur at low energy and ultrafast frequencies (THz and above) thus leading to competitive advantages in information processing design as compared to standard CMOS technology\cite{Jungwirth2016, Baltz2018}.
The basic concept of wave transmission and reflection phenomena is key to understanding how to optimize spin wave transport in these systems.
Recently, researchers established non-volatile control of thermal magnon transport in BFO using electric fields \cite{Parsonet2022}.
Their work demonstrates that the $109^\circ$ FE DWs act as a barrier to spin transport across a length-scale comprising many 100s of nm and dampen the detected magnon signal useful for the device.
We will illustrate the usefulness of our approach by showing how it can enable a mesoscopic simulation of this situation
We consider the two of the commonly observed DWs in BFO experiments, the $109^\circ$ 2/1 and $71^\circ$ 1/1 (100)-oriented boundaries\cite{Heron2014, Johann2012, Parsonet2022}.
The reader is referred to Table \ref{tab:FEdw1} and the previous section for the initial conditions of the order parameters.
There is a large relative difference between the lattice and spin DW energies.
This suggests that any application of an external magnetic field $\mathbf{H}_\mathrm{appl}$ should not appreciably influence the $\mathbf{P}$ and $\mathbf{A}$ subsystem.
Therefore, we fix in time the structural order parameters in this section.
We couple $\mathbf{H}_\mathrm{appl}$ to act on the net magnetization through the Zeeman free energy density,
\begin{equation}\label{eqn:Zeeman}
\begin{aligned}
    f_\mathrm{Zeeman} = -\mathbf{m}\cdot\mathbf{H}_\mathrm{appl}
\end{aligned}
\end{equation}
and add it to the total free energy of the spin configuration.
In order to perturb the system, we consider gaussian spin wave beams generated by a field of the form \cite{Gruszecki2015},
\begin{equation}\label{eqn:perturb}
\begin{aligned}
    \mathbf{H}_\mathrm{appl} &= H_0 \, \mathrm{sinc}[k_0 (x-x_0)] \, e^{-p_0(x-x_0)^2} \\ \nonumber
    &\times \mathrm{sinc}[\omega_0 (t-t_0)]\,\hat{\mathbf{h}}
\end{aligned}
\end{equation}
where field amplitude $H_0 = 184$ kOe, excitation location $x_0$, gaussian intensity profile parameter $p_0 = 0.16$ $\mathrm{nm}^{-2}$, and $k_0 = 10$ $\mathrm{nm}^{-1}$ control the perturbation distribution in spacetime.
The director $\hat{\mathbf{h}}$ orients the magnetic field with respect to $\mathbf{m}$.
Finally, we cut-off the pulse at $t_0 = 1$ ps and excite the spin waves at a frequency $\omega_0$. 

Eq.~(\ref{eqn:LLG_LLB}) is evolved with $\alpha = 0$ and Eq~(\ref{eqn:perturb}).
We enforce periodicity in our computational volume along the $x, y, z$ for the $\mathbf{m}_1$ and $\mathbf{m}_2$ variables.
The time-integration of Eq.~(\ref{eqn:LLG_LLB}) is set for $dt < 2$ fs time steps to ensure numerical convergence for the fast AFM dynamics in the system.
We verify that our calculations are in the linear limit by adjusting the $H_0$ and determine that the perturbed amplitudes of $\mathbf{m}_\eta$ scale linearly.
Finally, we monitor the system total free energy $F_\mathrm{sp} + F_\mathrm{MP}$ and $|\mathbf{m}_\eta|$ (via the LLB term) and verify that they are constant to within floating point accuracy for all time in our $\alpha = 0$ simulation.
In Fig.~\ref{fig:spin_pulse}(a), we track the \emph{excess} free energy density $f_\mathrm{exc}(t,x) = f_\mathrm{mag}(t,x) - f_\mathrm{mag}(t = 0,x)$.
Therefore, $f_\mathrm{exc}$ corresponds to a small energy that is injected into our computational volume by the spin excitation at time $t$.
A few snapshots of the $f_\mathrm{exc}(t,x)$ due to the propagating wavefront (at two different $\omega_0$) are presented in Fig.~\ref{fig:spin_pulse}(a) in sequential panels from top to bottom for $t = 4.5, 17.1, 24.6, 27.1$, and $34.5$ ps.
Here in panel (a), the DW is marked at $x_\mathrm{DW} = 22$ nm and is impacted by the spin wave at around $t = 24.6$ ps.
The excess energy density loss after the wavefront travels through the DW can be calculated by numerically time integrating $f_\mathrm{exc}(t,x)$ at distances of $\Delta x = 7$ nm left and right from the DW plane located at $x_\mathrm{DW}$.

We then compute their ratio $R$,
\begin{equation}\label{eqn:rec}
\begin{aligned}
    R = \frac{\int f_\mathrm{exc}(t, x_\mathrm{DW}+\Delta x) dt - \int f_\mathrm{exc}(t, x_\mathrm{DW}-\Delta x) dt}{\int f_\mathrm{exc}(t, x_\mathrm{DW}+\Delta x) dt}
\end{aligned}
\end{equation}
to determine which percentage of the excess energy due to the incoming wave is reflected or absorbed by the DW, i.e., the degree of rectification.
We see in Fig.~\ref{fig:spin_pulse}(b) that $R$ varies substantially across several decades in frequency with an asymptote for low frequencies corresponding to about 35 $\%$ and 50 $\%$ rectification for the 2/1 and 1/1 walls respectively.
The relative difference between rectification arises from the excitation of the DW region by the spin wave (seen in Fig. \ref{fig:spin_pulse}(a) for $t > 24.6$ ps).
To verify this, we track the time integrated $f_\mathrm{exc}$ at the DW revealing almost all of the excess energy is absorbed by the DW.
In this analysis, we find that only a small portion of $f_\mathrm{exc}$ due to this spin wave is reflected (not shown).
When the frequency is increased, a maximum in DW excess energy absorption in $R$ is acquired around 1-2 THz before $R$ abruptly decreases indicating that the DW becomes more transparent to the spin wave.
Similar frequency-dependent transmission ratios have been reported in the literature for noncollinear AFMs\cite{Rodrigues2021}.
We should mention that we did not find any meaningful influence of $\hat{\mathbf{h}}$ or $k_0$ on our results except changing the relative rectification between the two types of walls, but a more detailed study of the parameters is warranted.
Finally, we should comment on how this agrees with experimental observations.
In the work of Parsonet \emph{et al}\cite{Parsonet2022}, the propagating thermal magnon signal inferred from the inverse spin Hall effect was seen to decay exponentially as a function of distance from the source.
This was postulated as due to the 2/1 (100)-oriented DWs in the system acting as a barrier to spin currents with $\mathbf{k}||[100]$ whose number increased upon electrode separation; we can conclude that our results support this conclusion qualitatively.
It remains to be seen if domain engineering techniques guided by similar calculations could possibly help control the rectification that impedes efficient control of magnon signals in ME spintronics.
Also, we should mention that as opposed to our approach detailed in this section, in principle $\mathbf{P}$ and $\mathbf{A}$ could vary in time. This would lead to electromagnonic effects localized to the DWs upon excitation of inhomogeneous magnetic and/or electric fields (as highlighted in recent Refs. [\citen{Sayedaghaee2022a,Sayedaghaee2022b}]). 
It is possible that the coupled vibrations of the structural order could further influence the spin transport.
However, this is a outside the scope of this work and warrants future studies.
\subsection{Magnetoelectric switching of the AFM order}\label{sec:MEsw}

\begin{figure*}\centering
\includegraphics[scale=0.285]{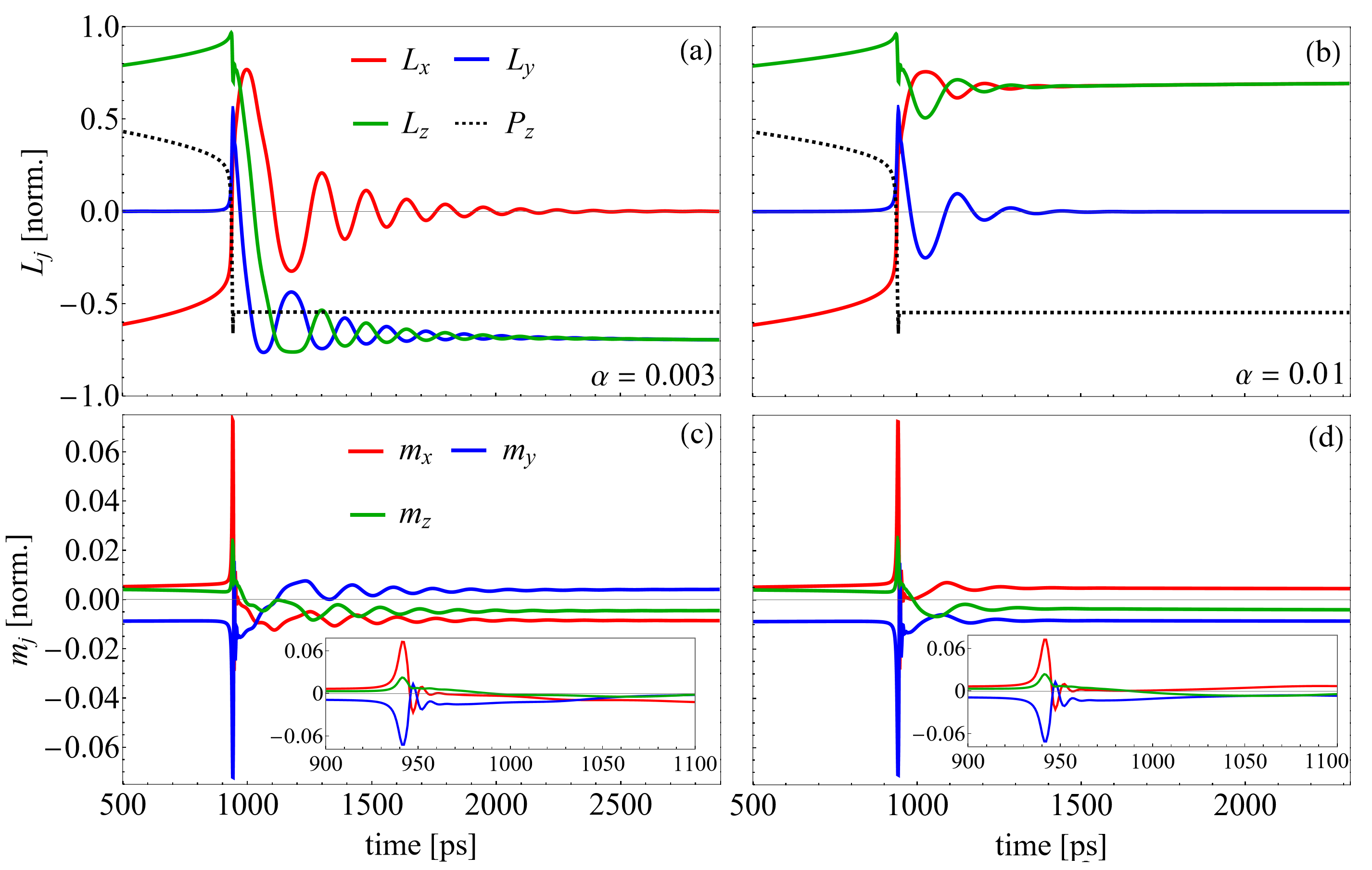}
\caption{\label{fig:MEsw} A switching event corresponding to a $71^\circ$ rotation of $\mathbf{P}$ (shown in the black dashed line) by application of an $\mathbf{E}$ with $\omega = 600$ MHz and $E_0  = 1800$ MV/cm. The N\'{e}el $\mathbf{L}$ components (normalized) are shown corresponding to $\alpha = 0.003$ and $\alpha = 0.01$ for (a) and (b). The switch (in $\mathbf{L})$ occurs from $[\bar{1}01]\to[0\bar{1}\bar{1}]$ (a) and $[\bar{1}01]\to[101]$ (b). The value of $\mathbf{m}$ settles into the minima corresponding to a $[1\bar{2}1]\to[\bar{2}1\bar{1}]$ and $[1\bar{2}1]\to[1\bar{2}\bar{1}]$ transitions respectively in (c) and (d). The insets in (c) and (d) show the similar ringdown time dependence near the $P_z$ switch (occuring at $t \approx 925$ ps).}
\end{figure*}

A considerable demand in AFM spintronics is to find an adequate approach to manipulate the magnetic order with external stimulii.
%
%
In the case of BFO, since this material displays an intrinsic electric dipole moment, it has been proposed to use an electric field to manipulate and control the magnetic texture.
The technological benefits to the prospect of electric field control of magnetism has been considered for some time \cite{Heron2014, Matsukura2015, Song2017, Liu2021, Parsonet2022}.
While low-frequency deterministic switching of $\mathbf{m}$ with an electric field has been experimentally demonstrated\cite{Heron2014}, the dynamical processes of the coupled polar-magnetic order is still a topic of research\cite{Liao2020a, Liao2020b}. 
We aim to highlight one such use of this modeling effort for the case of ME switching (i.e., using an electric field to switch $\mathbf{m}$).
We now consider a fully-dynamical simulation where all system variables $\{\mathbf{P},\mathbf{A},\mathbf{u},\mathbf{m}_1$, $\mathbf{m}_2\}$ depend on time.
As we are now interested in real dynamics, the time relaxation constants $\Gamma_P = 200$ $\mathrm{F}\mathrm{m}^{-1}\mathrm{s}^{-1}$ and $\Gamma_A = 83188$ $\mathrm{deg}^2\mathrm{m}^{3}\mathrm{J}^{-1} \mathrm{s}^{-1}$ in Eq.~(\ref{eqn:TDLG_P}) and (\ref{eqn:TDLG_A}) are taken from Ref. [\citen{PrivCommFedorova2023}].
%
%
For our switching simulations, our initial condition of the $\mathbf{P}\uparrow\uparrow\mathbf{A}$ system is along the $[111]$ direction and \emph{homogeneous}.
Since this is a homogeneous calculation, this can be considered the macrospin limit of Eq.~(\ref{eqn:LLG_LLB}). 
%
Since the dynamics of the AFM order are in general very fast (characteristic frequencies of $100$s of GHz to the THz regime)\cite{Jungwirth2016}, we introduce a time stepping constraint on the evolution of Eq.~(\ref{eqn:LLG_LLB}) for dt $ < 0.1$ ps to ensure numerical convergence.
There is no spin dissipation from conduction electrons in BFO due to its insulating nature.
Therefore, we choose $\alpha$ of order $10^{-3}$ which is a reasonable assumption for BFO \cite{Wang2012, Bhattacharjee2014} and other magnetic insulators \cite{Kapelrud2013, Shiino2016, Soumah2018, Chen2021}.

\begin{figure}\centering
\includegraphics[scale=0.42]{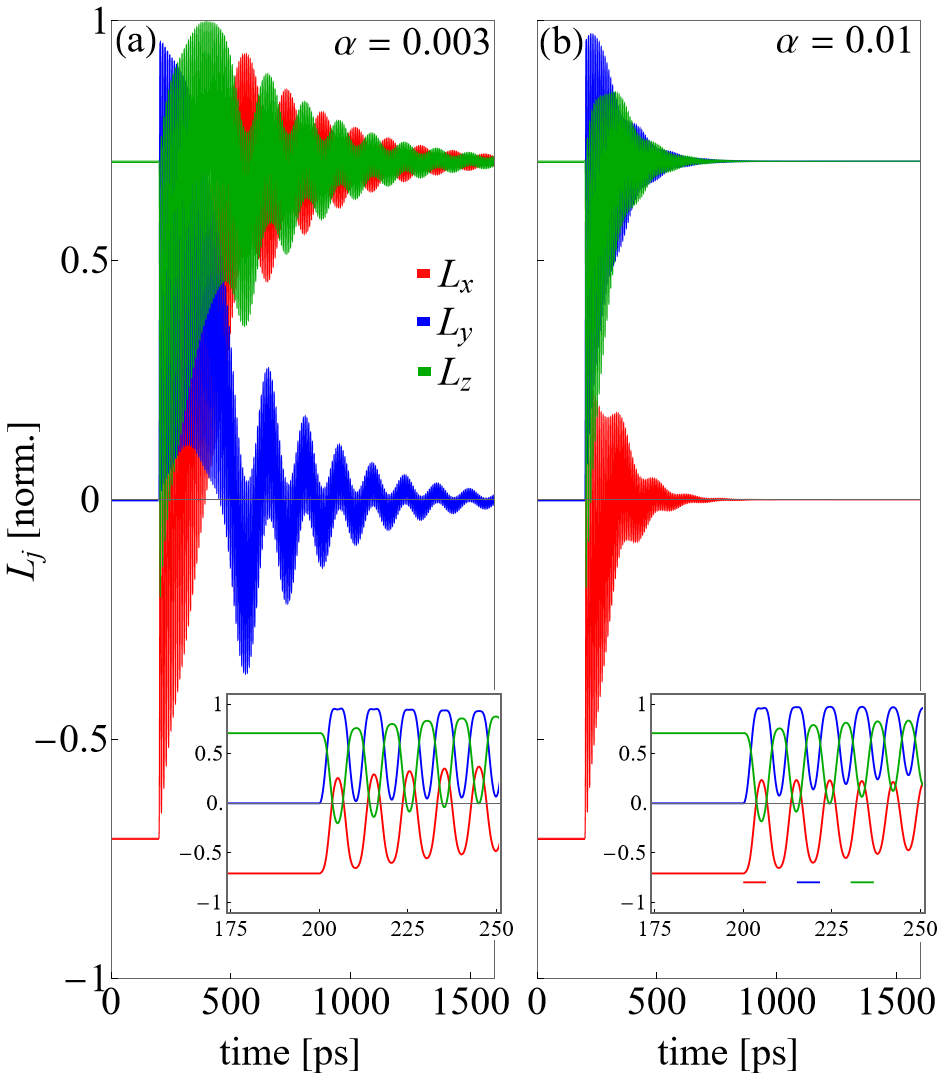}
\caption{\label{fig:MEswINST} Instantaneous switching of the N\'{e}el vector $\mathbf{L}$ for (a) $\alpha = 0.003$ and (b) $\alpha = 0.01$. The switch occurs abruptly at $t = 200$ ps causing the AFM order to rapidly oscillate. The initial state is $\mathbf{L}||[\bar{1}01]$ leading to final states of $\mathbf{L}||[101]$ and $\mathbf{L}||[011]$ in (a) and (b) respectively.}
\end{figure}
As one example to switch the $z$ component of $\mathbf{P}$, we choose our electric field $\mathbf{E}$ to be $\mathbf{E}(\omega) = \langle 0,0,E_0 \sin{\left(\omega t\right)}\rangle$ with $E_0 = -1800$ MV/m.
This is a large value compared to coercive fields of $E_c = 20-40$ MV/m observed in switching experiments of thin film BFO heterostructures\cite{Wang2003, Heron2014}.
However, it is well-known that the coercive field needed to fully switch components of $\mathbf{P}$ in perovskite FEs is intrinsically linked to the occurence of various phenomena\cite{Morozovska2005,Bratkovsky2000,Indergand2020, Gerra2005, Liu2016} that are not present in our homogeneous switching simulations. 
We select an \textbf{E} frequency of $\omega = 600$ MHz.
The field is abruptly turned off after $\mathbf{P}$ has switched in order to facilitate only one switching event for analysis.
The initial state is homogeneous $\mathbf{P}\uparrow\uparrow\mathbf{A}$ along [111] with $\mathbf{L}||[\bar{1}01]$ and $\mathbf{m}||[1\bar{1}1]$ as one of the possibilities listed in Table \ref{tab:magGs}.
In order to investigate if the ME switching has dependency on $\alpha$, we pick two different values $\alpha = 0.003$ and $\alpha = 0.01$ and evolve Eq.~(\ref{eqn:TDLG_P}) and (\ref{eqn:TDLG_A}) in the presence of the field.
Here we see in Fig.~\ref{fig:MEsw}(a) and (b), the application of $\mathbf{E}$ along the $z$ direction switches the $\mathbf{P}$ (and also $\mathbf{A}$, not shown) orientation to $[11\bar{1}]$ within $1000$ ps (dashed black line).
We use the notation $i\to f$ to denote initial states $i$ and final states $f$ for the $\{\mathbf{m},\mathbf{L}\}$ system.
The change of the energy surface through the magnetostructural coupling causes $\mathbf{L}$ to switch orientation from $[\bar{1}01]$ to $[0\bar{1}\bar{1}]$ in (a) and $[\bar{1}01]\to[101]$ in (b).
At the same time, the direction of $\mathbf{m}$ undergoes $[1\bar{2}1]\to[\bar{2}1\bar{1}]$ and $[1\bar{2}1]\to[1\bar{2}\bar{1}]$ transitions in (c) and (d).
When one compares the dynamics between the left and right panels of Fig. \ref{fig:MEsw}, it is evident that the choice of $\alpha$ influences the final $\{\mathbf{m},\mathbf{L}\}$ state despite having nearly identical ringdown patterns at the temporal vicinity of the $\mathbf{P}$ switch shown in the insets of (c) and (d).
Shortly after the switch ($t > 1000$ ps), the magnetization evolves differently, due to different maximum amplitudes, leading to transition pathways which overcome different energy barriers.
We also consider an \emph{instantaneous} limit of the switching process where the $P_z$ is switched immediately.
In Fig. \ref{fig:MEswINST}(a) and (b) which correspond to the same $\alpha$ values as in \ref{fig:MEsw}(a) and (b), the switch is set to occur at $t = 200$ ps (shown in the insets).
The relaxation of Eq.~(\ref{eqn:LLG_LLB}) with the damping set to $\alpha = 0.003$ and $0.01$ creates many oscillations with a characteristic ringdown frequency of approximately $127$ GHz.
We find that indeed, the same situation happens presented in Fig. \ref{fig:MEswINST}(a) and (b) as in Fig. \ref{fig:MEsw}(a) and (b) with the final states of $\mathbf{L}$ determined by its initial orientation and the final configuration of $\mathbf{P}$.
The vector $\mathbf{m}$ (not shown) has trajectories $[1\bar{2}1]\to[\bar{2}1\bar{1}]$ and $[1\bar{2}1]\to[1\bar{2}\bar{1}]$ in Fig.~\ref{fig:MEswINST}(a) and (b) respectively.
In the simulations corresponding to Fig. \ref{fig:MEsw}, the switching of $\mathbf{m}$ occurs in about a 200 ps time window, whereas with the instantaneous calculation, the switching pathway requires at least 1 ns to ring down $\{\mathbf{m},\mathbf{L}\}$ with realistic material values of $\alpha = 0.003$. 
This is far above the theoretical switching limit of 30 ps proposed by Liao and co-workers\cite{Liao2020a, Liao2020b} who also utilized a LLG model for the AFM order coupled to a Landau-type parameterization.
We stress that both of these numerical simulations are exercises for illustrative purposes and are simplified versions of the dynamic processes that would happen in an experiment.
Our calculations already suggest two things: (1) the Gilbert damping $\alpha$ controls the maximum amplitude of the oscillations and thus the final state, hence it needs to be understood in BFO to have a repeatable effect and that (2) the dynamics of the structural switching does not seem to be essential in controlling the switching pathway (i.e. comparing the explicit time dependent $\mathbf{E}$ calculations vs, the instantaneous $\mathbf{P}$-$\mathbf{A}$ switches).
A more detailed investigation remains for the future.

\section{Conclusions and outlook}
We have presented a continuum model for BFO able to treat the polar, octahedral tilt, spontaneous strain, and the AFM order in a single calculation.
This model is built upon micromagnetic and FE phase field approximations to the system order parameters.
Our model is benchmarked against the known behavior in this material - specifically, we have parameterized the FE DW profiles along with their spontaneous strain fields obtaining an energy hierarchy of possible states in agreement with DFT calculations\cite{Dieguez2013}.
We also provide simulations of $\{\mathbf{L},\mathbf{m}\}$ in the presence of low energy FE DWs revealing delicate features in the angular quantities characterizing the canted magnetism.
Next, we illustrated the usefulness of the model with two simple applications i) AFM spin waves traversing the multiferroic domain boundary highlighting a rectifying nature in qualitative agreement with recent experiments \cite{Parsonet2022} and ii) fully-dynamical ME switching in real-time which, interestingly, reveals a sensitivity of
switching pathways on the Gilbert damping.

There are many other phenomena in BFO that could be built upon our model.
As is often discussed in the literature regarding BFO, is the appearance of a long-period spin cycloid \cite{Agbelele2017, Burns2020} in which the proposed origin is underpinned by an asymmetric spin-current mechanism\cite{Gareeva2013, Popkov2016, Xu2021, Meyer2022} which is necessary to stabilize these patterns.
While the results of this paper are for the weakly non-collinear AFM order, one can appreciate that the presence of the spin cycloid might affect the outcome of our illustrative examples.
We also emphasize that the DMI expression in Eq.~(\ref{eqn:dmi}) is \emph{isotropic} and that the application of strain should break the symmetry which can lead to different AFM sublattice ordering as detailed in Ref. [\citen{Dixit2015}].
In principle, both the spin-flexoelectric (spin cycloid) and magnetoelastic (epitaxial strain) contributions could influence the antisymmetric exchange leading to drastically altered magnetization textures in the simulations.
In general also, this type of multiferroic modeling could be extended to other noncollinear antiferromagnets such as those where electric fields have been shown to manipulate the magnetic state despite lack of spontaneous FE order\cite{He2010, Kosub2017}.
The model is built within the \textsc{Ferret} \cite{Mangeri2017} module atop the open-source Multiphysics Object Oriented Simulation Environment (MOOSE) framework \cite{permann2020moose}. 
As a nod to open-science, we provide representative examples for all of the results in this paper to be hosted on a GitHub website\cite{FerretLink}.
\textsc{Ferret} is part of a forward-integrating toolset called the Continuous Integration, Verification, Enhancement, and Testing (CIVET) \cite{slaughter2021continuous} utility which preserves reproducibility of our results by ensuring underlying code changes to the MOOSE software stack do not break the module.
The sets of governing equations and energy terms in this paper, which are applicable in 3D and for any geometry, are available and documented as \texttt{C++} objects within the open-source software repository.
While our modeling effort is certainly not exhaustive, we believe it will be a useful platform for development of continuum simulations of BFO and other multiferroics in length and time-scales are not accessible by atomistic methodologies.
\begin{acknowledgments}
The authors thank Natalya Fedorova for valuable input. J. M. has received funding from the European Union’s Horizon 2020 research and innovation programme under the Marie Sk\l{}odowska-Curie grant agreement SCALES - 897614. Work by O.H. supported by Basic Energy Sciences Division of the Department of Energy. D.R.R. acknowledges funding from the Ministerio dell'Università e della Ricerca, Decreto Ministeriale n. 1062 del 10/08/2021 (PON Ricerca e Innovazione).

\end{acknowledgments}

\appendix

\section*{Appendix}

In micromagnetic simulations in which thermal fluctuations are not included (so-called athermal simulations), the magnetization density must remain constant in magnitude, thus preserving the unit norm of the magnetization director $\hat{\mathbf m}={\mathbf M}/|{\mathbf M}|$. In finite-difference codes on regular meshes, such as MuMax3\cite{vansteenkiste}, enforcing this constraint is simple: each simulation cell $k$ contains one unit vector $\hat{\mathbf{m}}_k$ that can simply be renormalized after, {\em e.g.}, each time step. In contrast, in weak-formulation FEM codes the continuum magnetic degrees of freedom are approximated by shape functions on irregular mesh cells and integrated over, and normalization of ${\mathbf m}({\mathbf r})$ is not as easily interpreted or made meaningful as in finite-difference codes. 
One can overcome this problem, for example using, a representation of ${\mathbf m}({\mathbf r})$ in spherical coordinates\cite{Yi2014}, but numerical solutions of the equations of motion can become unstable leading to serious convergence issues. Another possibility is to introduce the constraint through a Lagrangian multiplier or using special shape functions on the tangent plane of the magnetization director vector field\cite{alouges2006convergence,szambolics}. We chose a different path, which is physically grounded in the Landau-Lifshitz-Bloch (LLB) formulation\cite{garanin1997,Garanin2004,evans2012}.
The key point in the LLB formulation is that longitudinal fluctuations in the magnetization director are allowed, but countered by a penalty for deviations away from the thermodynamic average of the magnitude $m(T)$ at a temperature $T$. The longitudinal fluctuations add a term to the equation of motion that is given by 
\begin{equation}
    \frac{\gamma\alpha_\parallel}{(1+\alpha^2)m({\mathbf r})^2}\left[\hat{\mathbf{m}}({\mathbf r})\cdot\left(\mathbf{H}_{eff}+\mathbf{\zeta}_\parallel\right)
    \right]\hat{\mathbf{m}}({\mathbf r}).
\end{equation}
where $\hat{\mathbf{m}}({\mathbf r})$ is the local magnetization director with an  equilibrium value $m_e(T)$ that depends on temperature, $\zeta_\parallel$ is a thermal field, and $\alpha$ is the usual dimensionless Gilbert damping.  
The longitudinal damping $\alpha_\parallel$ depends on $T$ through
\begin{equation}
    \alpha_\parallel=\alpha\frac{2T}{3T^{\rm MFA}_c}
\end{equation}
with $T^{\rm MFA}_c$ the mean-field Curie temperature, 
and the effective field $\mathbf{H}_\mathrm{eff}$ includes the longitudinal susceptibility $\chi_\parallel$,
\begin{eqnarray}
    \mathbf{H}_{\rm eff} & = &\mathbf{H}_{\rm ext}+\mathbf{H}_{\rm ani}+\mathbf{H}_{\rm ex}+
    \frac{1}{2\chi_\parallel}\left(1-\frac{m_i^2}{m_e^2}\right)\hat m_i\nonumber\\
   &  = &\mathbf{H}_0+\frac{1}{2\chi_\parallel}\left(1-\frac{m_i^2}{m_e^2}\right)\hat m_i.
\end{eqnarray}
Here $\mathbf{H}_{\rm ext}$, $\mathbf{H}_{\rm ani}$, and $\mathbf{H}_{\rm ex}$ are the usual external, anisotropy, and exchange fields. Ignoring the thermal field, the contribution to $d\hat{\mathbf{m}}/dt$ is then
\begin{equation}
    \frac{\gamma\alpha_\parallel}{(1+\alpha^2)m_i^2}
    \left[
    \hat m_i\cdot\left(\mathbf{H}_0+\frac{1}{2\chi_\parallel}(1-\frac{m_i^2}{m_e^2})\hat m_i\right) \right]\hat m_i.
\end{equation}
At $T=0$ with $m_e^2=1$ we can simplify the last term in the above equation to get
\begin{equation}
    \frac{\gamma\tilde\alpha_\parallel}{(1+\alpha^2)}\left(1-m^2\right)m^2 \hat{\mathbf{m}}
\end{equation}
where $\tilde\alpha_\parallel=\alpha_\parallel\mu_0/(2\chi_\parallel)$ now has the unit of a magnetic field that drives the longitudinal relaxation. The contribution to the time evolution of $\hat{\mathbf{m}}$ due to longitudinal relaxation is then
\begin{equation}
    -\frac{\gamma\tilde\alpha_\parallel}{(1+\alpha^2)}(m^2-1)m^2\hat{\mathbf{m}}
    +\frac{2 \chi_\parallel\gamma\tilde\alpha_\parallel}{(1+\alpha^2)m^2}(\hat{\mathbf{m}}\cdot\mathbf{H}_0)\hat{\mathbf{m}}.
    \label{eq:app_eq1}
\end{equation}
One can show that in the limit of low $T$, much lower than relevant Curie temperatures, the second term in Eq.~(\ref{eq:app_eq1}) can be ignored. 
In this case, the LLB-like addition to the equations of motion is simply 
\begin{equation}
    \frac{\gamma\tilde\alpha_\parallel}{(1+\alpha^2)}
    \left[m^2-1\right]m^2\hat{\mathbf{m}},
\end{equation}
where $\tilde\alpha_\parallel$ has the dimension of a field.

\bibliography{apssamp}

\end{document}